\pdfoutput=1

\documentclass[10pt, conference, letterpaper]{IEEEtran}
\usepackage{graphicx}
\usepackage[caption=false]{subfig}
\usepackage{amsmath}
\usepackage{amsfonts}
\usepackage{amssymb}
\usepackage{color}
\usepackage{tabularx}
\usepackage{multirow}
\usepackage{cite}

\usepackage[ruled]{algorithm2e}

\SetAlFnt{\small}
\SetKwRepeat{Do}{do}{while}

\makeatletter  \def\@eqnnum{{\normalsize \normalcolor (\theequation)}}  
\makeatother

\newcommand{\sunghan}[1]{{#1}}

\usepackage{setspace}

\usepackage{csquotes}
\ifCLASSINFOpdf
\else
\fi
\begin{document}
\title{SC-Share: Performance Driven Resource Sharing Markets for the Small Cloud}

\author{\IEEEauthorblockN{Sung-Han Lin, Ranjan Pal, Marco Paolieri, Leana Golubchik}
\IEEEauthorblockA{Department of Computer Science, University of Southern California}
\IEEEauthorblockA{\{\emph{sunghan, rpal, paolieri, leana}\}@usc.edu}
}


\maketitle

{\setstretch{1}
\begin{abstract}
Small-scale clouds (SCs) often suffer from resource under-provisioning during peak demand, leading to inability to satisfy service level agreements (SLAs) and consequent loss of customers. One approach to address this problem is for a set of autonomous SCs to share resources among themselves in a cost-induced cooperative fashion, thereby increasing their individual capacities (when needed) without having to significantly invest in more resources. A central problem (in this context) is how to properly share resources (for a price) to achieve profitable service while maintaining customer SLAs. To address this problem, in this paper, we propose the \emph{SC-Share} framework that utilizes two interacting models: (i) a \emph{stochastic performance model} that estimates the achieved performance characteristics under given SLA requirements, and (ii) a \emph{market-based game-theoretic model} that (as shown empirically) converges to efficient resource sharing decisions at market equilibrium. Our results include extensive evaluations that illustrate the utility of the proposed framework.
\end{abstract}}

\begin{IEEEkeywords}
data centers; small cloud; performance; markets
\end{IEEEkeywords}

% \fi
\IEEEpeerreviewmaketitle

\section{Introduction} \label{Sec:intro}

Infrastructure-as-a-Service is quickly becoming a ubiquitous model for providing elastic compute capacity to customers who can access resources in a pay-as-you-go manner without long-term commitments, with rapid scaling (up or down) as needed \cite{armbrust2010view}.
Cloud service providers (Amazon AWS \cite{aws}, Google Compute Engine \cite{computeengine}, and Microsoft Azure \cite{azure}) allow customers to quickly deploy their services without a large initial infrastructure investment.  
%
%Such public cloud providers invest in large-scale data centers, that are typically over-provisioned in order to be able to respond to bursty workloads during peak hours. 

\noindent\textbf{\emph{Proliferation of smaller-scale clouds}}.
However, there are some non-trivial concerns in obtaining services from large-scale public clouds, including % loss of data {\em privacy}, 
{\em cost\/} and {\em complexity\/}.
%
%Other concerns include of building services. \ranjan{To expand on these concerns,}
%
Massive cloud environments can be costly and inefficient for some customers, e.g., Blippex \cite{blippex}, thus resulting in more and more customers building their own smaller-scale clouds (SCs) \cite{cloudservey} for better control of resource usage; e.g., it is hard to guarantee network performance in large-scale public clouds due to their multi-tenant environments \cite{mogul2012we}.
%, while smaller-scale cloud providers can offer resources at more competitive prices due to their lower initial investment and maintenance fees, and simpler multiplexing environments; e.g., DigitalOcean \cite{digitalocean} offers higher specification VMs than Amazon EC2 at similar or lower prices.
%
Moreover, smaller-scale providers exhibit a greater flexibility in customizing services for their users, while large-scale public providers 
%have greater limitations, partly due to their large multi-tenant environments \cite{blippex}
minimize their management overhead by simplifying their services; e.g., Linode \cite{linode} distinguishes itself by providing clients with easier and more flexible service customization.
%
%\ranjan{you have to mention something about privacy with citations here to close the loop in this paragraph. in the other journal paper I submitted on small clouds, you will find some text and citation} 
%Thus, many smaller-scale clouds are available \cite{spotcloud}. \ranjan{This is not the right way to lead to a conclusion, need to reframe the previous sentence.}
%
The use of SCs is one approach to resolving cost and complexity issues.

Despite the potential of SCs, they are likely to suffer from resource under-provisioning during peak demand, which can lead to inability to satisfy service level agreements (SLAs) and consequent loss of customers.
SLAs come in many forms, such as the average or maximum waiting time before being served, the probability of requests being rejected, and the amount of resources that each request can obtain. 
%(The service time might not be a factor here because the duration of time mainly depends on the complexity of requests.) \ranjan{This is not the place to say whether service time will be a factor}
%
In order not to resort, similarly to large-scale providers, to resource over-provisioning, with all its disadvantages, one approach to realizing the benefits of SCs is to adopt hybrid architectures \cite{zhang2014proactive, shifrin2013optimal}, that allow private clouds (or small cloud providers) to outsource their requests to larger-scale public providers. 
However, the use of public clouds can potentially be costly for the small-scale provider.
%, and (b) still leave the small-scale provider open to privacy concerns.

\noindent\textbf{\emph{Motivation.}}
An emerging approach to solving the under-provisioning problem is for SCs to share their resources in a \emph{federated cloud environment} \cite{babaoglu2012design,zhuang2014decentralizing,toosi2011resource,mashayekhy2015cloud,chen2017workload,goiri2012economic,samaan2014novel,wangenabling,hassan2014cooperative,hadji2015mathematical,wen2016cost}, \sunghan{thus (effectively) increasing their individual capacities (when needed) without having to significantly 
invest in more resources, e.g., this can be helpful when the SCs do not experience peak workloads at the same time.
Earlier efforts \cite{goiri2012economic,hassan2014cooperative} characterize the benefits of cloud federations, while \cite{samaan2014novel} also demonstrates that the uncertainty in meeting SLAs can be an incentive enabling sharing of resources among clouds.
Moreover, the ability of utilize to multiple SCs can avoid single points of failure: when one SC suffers an outage, others can be accessed to rent VMs.
For instance, on February 28th, 2017, AWS suffered a five-hour outage in the US, causing an estimated damage of $\$150,000,000$ to $S\&P 500$ companies.}
%
%\footnote{
%	Since one benefit of small-scale clouds is management of privacy concerns, one should consider that in determining with whom to cooperate; however, this is outside the scope of this paper.}.
% 
%\ranjan{Sung-Han this last sentence is not at all clear. We need to work on this}
%
%In this work, we refer to SCs sharing their resources with each other as a {\em cooperative}.
%
%\cite{hassan2014cooperative} also studies the motivation for cloud providers to participate in the federation. 
%
%Authors of \cite{wangenabling} present SpotCloud \cite{spotcloud}, a cooperative system that helps providers selling idle resources to other providers or end-users at specified prices, and they analyze pricing models that incentivize providers to contribute their resources.
%
%Other efforts in cloud sharing domain  
%, and focus on designing efficient sharing mechanisms .

However, many of these efforts assume the existence of the cloud federation and largely focus on designing sharing policies in order to maximize the profit of individual SCs \cite{zhuang2014decentralizing,toosi2011resource,hadji2015mathematical,wen2016cost}.
For example, \cite{toosi2011resource} proposes a strategy to terminate less profitable spot instances, in order to accommodate more profitable on-demand VM requests.
%
%Moreover, most of them focus on the performance of the cooperative as whole, rather than on the potential benefits (in terms of profit) and cost (in terms of performance degradation) to individual SCs, which are significant contributing factors to incentivizing SCs to participate in the cooperative. 
%
\sunghan{Moreover, most works do not consider the trade-off between economical benefits (in terms of profit) and  performance degradation for individual SCs, which is a significant factor in incentivizing SCs to participate in the cloud federation.
Without the analysis of performance degradation due to resource sharing, the feasibility of a federation can be questioned.
\cite{mashayekhy2015cloud} studies a federation formation game among cloud providers based on revenue. However, it only considers a special scenario where all cloud providers share all their resources with others.}
Thus, this work focuses on a fundamental, unanswered question of ``{\em how each SCs should share resources to be profitable without violating customer SLAs, while also motivating other SCs to join the federation}.''

\begin{figure}
	\centering
	\includegraphics[width=0.85\linewidth]{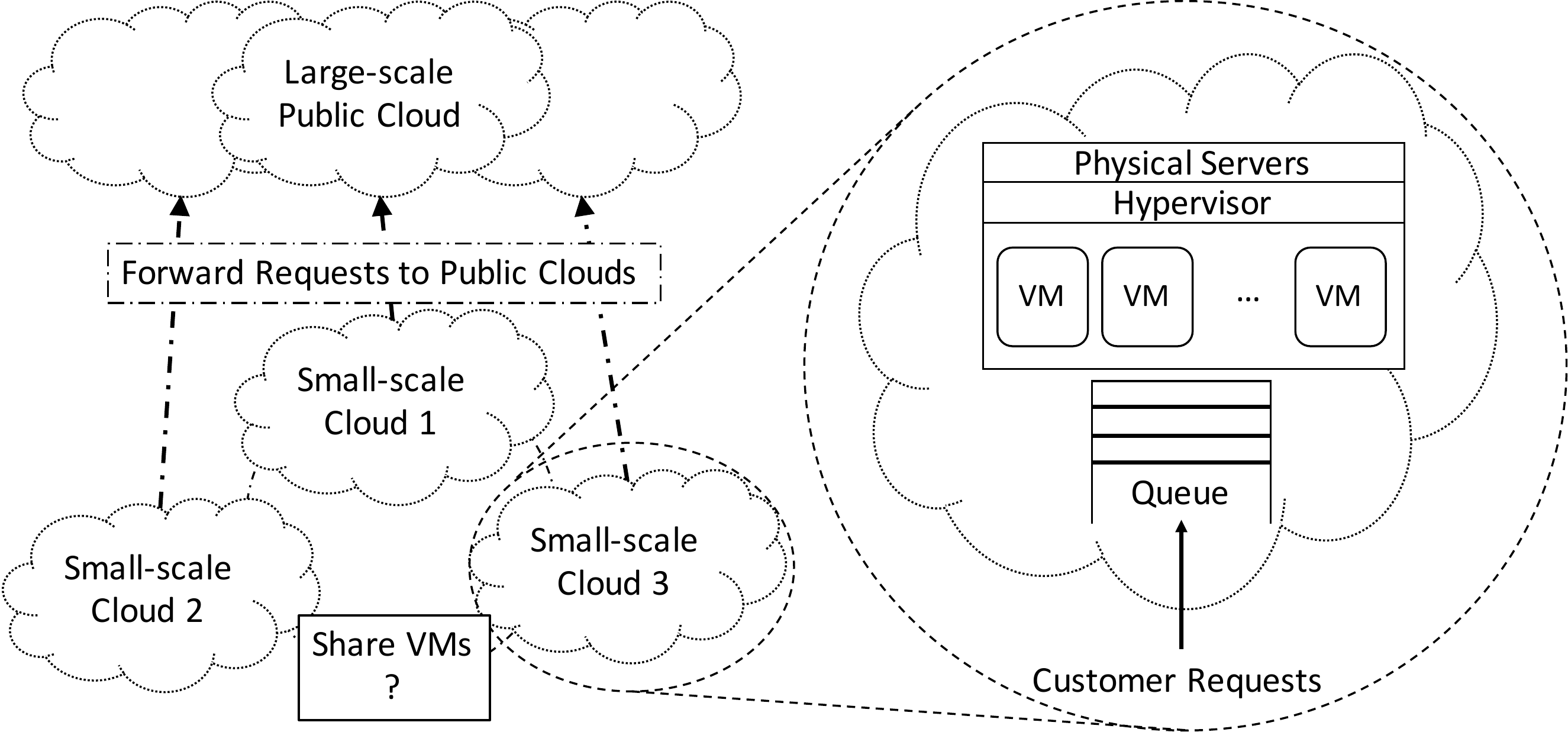}
	\caption{System Overview}
	\vspace{-1em}
	\label{fig:cloud_share_overview}
\end{figure}

\noindent\textbf{\emph{Problem description}}.
We consider an environment with multiple SCs; an example with $3$ SCs is depicted in Fig.\ \ref{fig:cloud_share_overview}.
In this work, we also refer to SCs sharing resources with each other as a {\em federation}.
%
%Each small-scale cloud has its own SLAs with its customers: the maximum waiting time before getting served, which are the important factors for the service fee for customers. (The shorter waiting time, the more pricey.)
Each SC has its own SLAs with its customers: {\em the maximum waiting time before service of a request is initiated}.
%
%In order to serve requests by the defined waiting time, 
To satisfy SLAs, SCs use public clouds as a ``backup'', i.e., buy needed resources on-demand from large-scale public clouds, when in danger of not being able to meet SLAs.  
If such SCs form a federation, in the event that an SC runs out of its resources, it can first use shared resources from other SCs for a price lower than the price of using public clouds. 
%(All participants are not expected to experience peak loads simultaneously.)
The amount of shared resources directly affects how much workload the federation is able to handle, which in turn affects the profit each SC is able to achieve.
In this sharing scenario, an important question is: {\em should SCs participate in the federation? If yes, how much should each SC share}? 
If an SC is too generous (i.e., shares too many of its resources), then it may be in danger of not being able to serve its own workload, resulting in more requests being forwarded to public clouds thereby reducing profit margins.
As a result, an SC should determine the amount of resources shared based on the price of selling and buying %its resources and buying participants' 
resources, i.e., the net profit compared with the cost of using public clouds.
However, if an SC is too selfish, i.e., shares few of its resources for higher profit, then either it may get removed from the federation for not being a useful contributor, or the federation may fall apart if most/all SCs tend towards selfish behavior.
Thus, another critical question that needs to be addressed is: {\em what price can make each SC share a reasonable amount of resources so that all SCs will participate in the federation?}
%\ranjan{Need a heading for the following text}

\noindent\textbf{\textit{Challenges and contributions.}}
To answer these questions, we make the following contributions:
\begin{enumerate}
\item \textbf{{\em Performance-dependent cost function:}}
Operating costs of an SC depend on the SLA with its customers and on the performance achieved inside the federation; in particular, we need to compute how frequently the SC will need to allocate external resources to satisfy SLAs (e.g., maximum waiting time), and whether it will be able to use resources of other SCs, or only those of public clouds.
In Sect.~\ref{Sec:initial_model}, we develop a \textit{detailed performance model} to compute such performance metrics for each SC. In turn, these metrics allow us to compute the operating cost of SCs (as defined in Sect.~\ref{Sec:cost}).
To address the high computational complexity of the detailed performance model (due to its large state space, which grows exponentially with the number of SCs), we develop an \textit{approximate performance model} (Sect.~\ref{Sec:approximate_model}). \sunghan{This model provides accurate estimates of the measures of interest, with linear complexity in the number of SCs, without requiring SCs to leak their sensitive information.}
% (which is crucial, given that SCs must take sharing decisions almost in real-time). 

\item \textbf{{\em Sharing market design:}}
The sharing mechanism should motivate SCs to participate, without significant oversight nor management, i.e., they should find an economic benefit in contributing resources to the federation.
We design a market-based model to determine the price charged within the federation for the use of shared resources.
The model is based on a non-cooperative, repeated game among SCs, each being selfish and trying to maximize its utility; as in real-world scenarios, SCs do not know the utility of other SCs, but they can compute (using our approximate performance model) the operating cost that they would incur for each possible sharing decision.
We determine {\em market equilibrium} conditions under which the federation is successful and {\em market efficiency} is achieved (Sect.~\ref{Sec:market_model}).

\item \textbf{{\em Experimental evaluation:}} In Sect.~\ref{Sec:eval}, we perform an extensive experimental evaluation to validate the accuracy of our approximate performance model with respect to simulation, and to verify the existence of market equilibria. Results highlight errors lower than 10\% for the performance metrics of interest; the proposed pricing model achieves market equilibria and good economic efficiency, successfully incentivizing SCs to stay in the federation.
\end{enumerate}
To the best of our knowledge, ours is the first work that models small-cloud federations as a holistic \emph{performance-driven market}, integrating engineering aspects (from a performance model) with economic ones (from a market model).
\section{System Description}
\noindent
In this section, we first describe the architecture of the SC federation, illustrated in Fig.~\ref{fig:cloud_share_overview}.
We then introduce a definition of operating costs of SCs.
Finally, we describe our sharing framework, which we call \textit{SC-Share}. 

\subsection{Architecture Description}

Each SC has a number of physical servers: through virtualization technology, physical resources (CPU, memory, storage) of SC~$i$ are packed into $N_i$ homogeneous virtual machines (VMs), which are the resource unit adopted in this work.
Customers request the allocation of individual VMs from SCs; the arrival process of VM requests at each SC~$i$ is modeled as a Poisson process with  rate $\lambda_i$.
The service time of each request at SC~$i$ (including the time elapsed from start of VM preparation until its release by the user) is modeled as an exponential random variable with rate $\mu_i$.
Each SC processes VM requests in FCFS order.
%
%When receiving a VM request, an SC identifies available resources among all its physical servers, and provides them to customers in the form of VMs.
%
%Due to resource limitations, each physical server can only host a limited number of VMs at any given time.
%
\sunghan{If physical servers do not have sufficient resources for a new VM, an SC can reject the request, queue it until more resources are available, or forward it to a public cloud (in a hybrid-cloud model).
%
%This SC can even forward queued requests to a large public cloud (e.g., Amazon), i.e., it will purchase resources from a public cloud, when it cannot satisfy its customers' QoS with its own resources.
In Sect.~\ref{Sec:discussion}, we discuss the details of these assumptions.}

In a federation with $K$ SCs (Fig.~\ref{fig:cloud_share_overview} depicts the case $K=3$), we consider the following general scenario:
when all VMs at an SC are fully occupied, its new VM requests are queued and can be served either by waiting for local resources to become available, or by purchasing resources from other SCs in the federation, or from a public cloud.
In order to participate in the federation, SC~$i$ must determine the maximum number of VMs $S_i$ to share with other SCs (at a given price) when idle VMs are available; i.e., at any time instant, the number of VMs shared by SC~$i$ is $I_{i}^{S_i} \le S_i$.
When all its VMs are occupied, SC~$i$ cannot terminate VMs serving requests of other SCs, but only stops accepting such requests until it is able to clear its own queue.
Each SC $i$ is required to maintain SLAs with its customers; we assume that this corresponds to a bound on the waiting time, i.e., a VM needs to be provided by SC~$i$ within $Q_i$ time units from its request.
If SC $i$ determines that it is not able to satisfy this SLA using resources of the federation, it forwards the request to a public cloud (e.g., Amazon AWS).

\subsection{Cost Metric Description} \label{Sec:cost}
SCs usually make large up-front investments in infrastructure, and continue to pay for maintenance costs (e.g., power supply and cooling costs).
In addition, SCs need to consider costs for forwarding requests to public clouds or for using resources in the federation, in order to satisfy customer SLAs.
We define a \emph{cost metric} to combine these costs with the revenue generated by VM requests from other SCs in the federation, and compute the net operating cost.

Let $I_i^{S_i}$ be a random variable representing the number of SC~$i$'s VMs per second used by other SCs when SC~$i$ shares up to $S_i$ VMs with the federation.
Let $O_i^{S_i}$ and $P_i^{S_i}$ be random variables representing the number VMs per second used by SC~$i$ from the federation and from a public cloud, respectively, to satisfy its SLAs.
The net cost for SC $i$ is then
\begin{equation}\label{Eq:cost}
C_i^{S_i} = \overline{P_i^{S_i}} \cdot C_i^{P} + (\overline{O_i^{S_i}} - \overline{I_i^{S_i}}) \cdot C_i^{G}\quad \forall i,
\end{equation}%
where $C^{P}_i$ and $C^{G}_i$ represent the cost of using a single VM from a public cloud and from other SCs,
%(how to determine $C^{P}_i$ and $C^{G}_i$ is out of the scope of this paper), 
respectively.
$\overline{P_i^{S_i}}$, $\overline{O_i^{S_i}}$, and $\overline{I_i^{S_i}}$ are the mean number of VMs per second used by SC $i$ from a public cloud, by SC~$i$ from other SCs, or by other SCs from SC~$i$, respectively.
\sunghan{Here, $\overline{P_i^{S_i}} \cdot C_i^{P}$ is the cost (penalty) for not serving requests locally, which drives SCs to participate in the federation and determines proper sharing decisions since we assume that $C^{P}_i > C^{G}_i$.}
{\em To reduce cost, by making appropriate sharing decisions, i.e., determining the number of VMs to share with others, %at any time instant, 
we need a performance model for each SC, in order to properly estimate $\overline{P_i^{S_i}}$, $\overline{O_i^{S_i}}$, and $\overline{I_i^{S_i}}$
(see Sect.~\ref{Sec:performance_model} for details).}
Unlike~\cite{toosi2011resource}, where cloud providers change VM prices based on system utilization, our model considers a fixed price $C^{G}_i$ for every VM.
%charge the use of VMs for the price 
%
Since VMs are homogeneous, we assume that $C_i^G = C_j^G\:\forall i,j=1,...,K$, but each SC can have a different $C_i^P$, depending on which public cloud it uses.
%
%These assumptions simplify our performance model, and where SCs allocate available VMs in the federation without preferences due to prices.
\sunghan{(This assumption is discussed in Sect.~\ref{Sec:discussion}.)}

Another incentive for participating in the federation is reducing power cost by forwarding VMs to other SCs when they offer VMs at cheaper prices than the cost of instantiating VMs in SC's own environment.
For instance, previous efforts \cite{goiri2012economic,kessaci2013pareto} study the sharing mechanisms for cloud providers to minimize their costs.
However, in this work, we only focus on the cost of additional resources required to satisfy customers' SLAs.
Extending the cost function to incorporate power consumption of executing VMs is a future direction.

\begin{figure}
	\centering
	\includegraphics[width=0.45\textwidth]{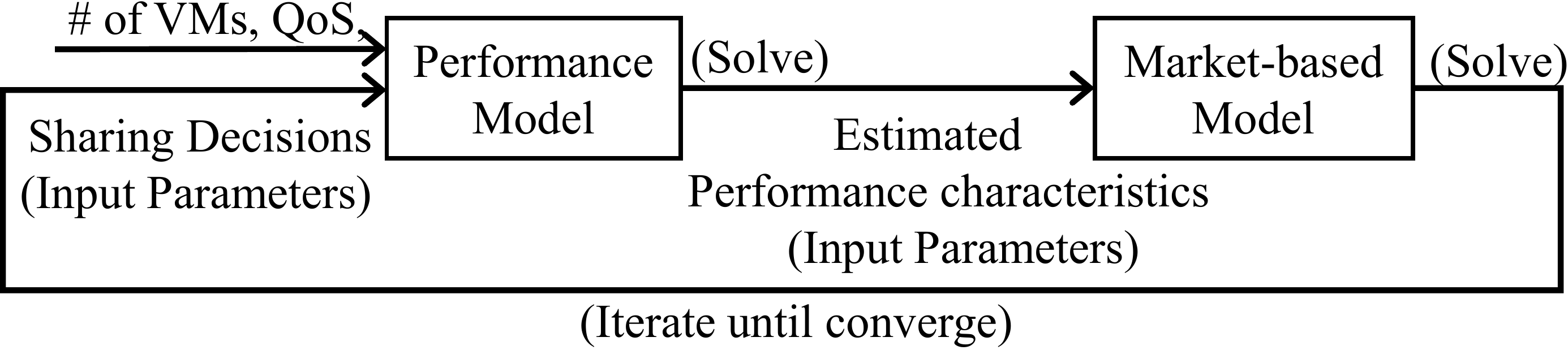} 
	\caption{Feedback between two models}
	\vspace{-1em}
	\label{fig:steps}
\end{figure}

\subsection{Cost Metric Evaluation Framework} \label{Sec:metric_evaluation}

In order to help SCs determine whether it is beneficial for them to participate in the federation and share their resources, we design a framework \emph{SC-Share} that allows each SC $i$ to determine the best value of $S_i$, in order to maintain its SLA~$Q_i$ and minimize the expected operating cost $C_i^{S_i}$.

The essence of SC cooperation in such a federation is the mutual agreement among individual SCs
% (in the presence of a regulator, e.g., federation among SCs) 
to share their resources (if idle) with other SCs experiencing peak workloads.\footnote{The issue of enforcing the agreement is beyond our scope here.} 
However, the amount of resources $S_i$ that each selfish but honest SC wants to share represents its strategic property that subsequently affects the cost metric $C_i^{S_i}$. 
Thus, in \emph{SC-Share}, we develop a market-based model to capture SC interactions in the federation via a market consisting of $K$ selfish SCs that interact strategically, and repeatedly over time, via a non-cooperative game to converge upon stable parameter values. 

However, a feedback loop exists between the performance model and the market model: sharing decisions $S_i\;\forall i$ are used by the performance model to compute $\overline{P_i^{S_i}}$, $\overline{O_i^{S_i}}$, and $\overline{I_i^{S_i}}$ and evaluate the cost metrics $C_i^{S_i}$ in Eq.~(\ref{Eq:cost}), which, in turn, determine the SC utility functions of the market model governing sharing decisions.
Therefore, in \emph{SC-Share}, we propose an iterative solution approach, as illustrated in Fig.~\ref{fig:steps}, involving these two models and their mutual feedback, to converge upon stable sharing decisions.

\section{Performance Model} \label{Sec:performance_model}

In this section, we propose a performance model for \emph{SC-Share} that is used to compute performance parameters required by the cost function of Eq. (\ref{Eq:cost}).

\subsection{SC without Sharing Resources} \label{Sec:forward_model}

We start with a degenerate case, where an SC does not participate in the federation and shares no VMs.
Based on SLA requirements, the SC will forward a request to public clouds if service cannot be started within $Q_i$ time units after its reception.
To compute the cost, we need to estimate the mean number of requests forwarded per second by SC~$i$, $\overline{P_i^{0}}$ (we denote it with ``$0$'' since no VMs are shared). %, when SC~$i$ does not participate in the cooperative and share no VM.

%
%Thus, in the hybrid cloud architecture, a request will be forwarded to public clouds if SC~$i$ determines that it cannot guarantee $Q_i$ for the request.
%
%If this forwarding probability is high, it will motivate SCs to participate in the cooperative.
%
%To this end, SC needs to have a model to  and the corresponding cost for forwarding requests to public clouds.

\begin{figure}
	\centering
	\includegraphics[width=0.40\textwidth]{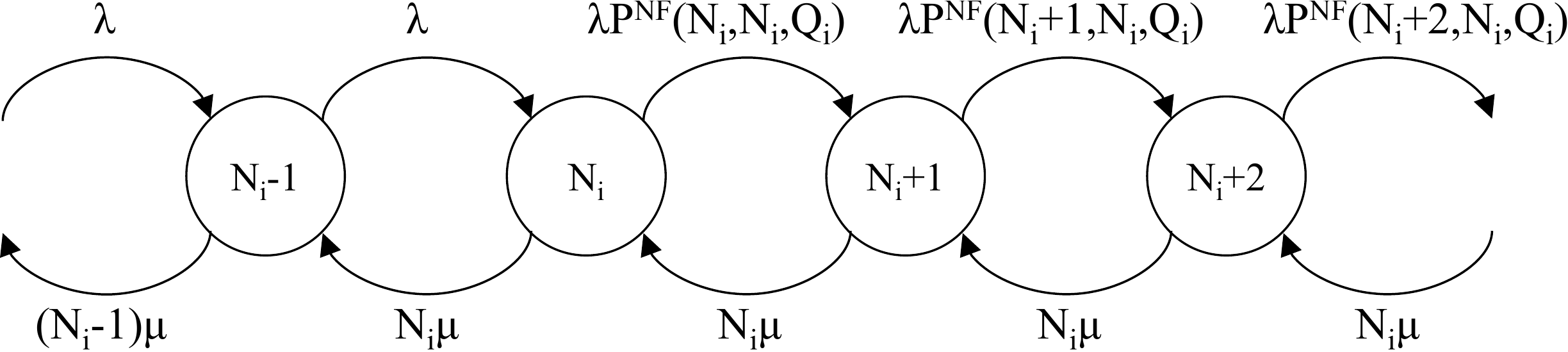} 
	\caption{A Markov model for forwarding}
	\vspace{-1.5em}
	\label{Fig:forward_public}
\end{figure}

To compute $\overline{P_i^{0}}$, we use a Markovian model, where the state represents the number of requests at SC~$i$, as illustrated in Fig.~\ref{Fig:forward_public}.
In this example, we assume that SC~$i$ has $N_i$ VMs and SLA $Q_i$ with its customers.
When at least one VM is idle, a new request can be served immediately.
%
%However, when all VMs are occupied, to capture the idea of forwarding requests, we make the arriving request queue in the system with a probability of serving this new arrival in $Q_i$ time based on the current number of queued requests.
However, when all VMs are busy, the probability that the new request is added to the queue of SC~$i$ (rather than forwarded to a public cloud) is equal to the probability that service will start in $Q_i$ time units, based on the current number of queued requests.
Let $q_i$ be the number of customers in SC~$i$ (i.e., $\max(0, q_i - N_i)$ customers are waiting in its queue) at the time of the request arrival.
Then, %given exponential inter-arrival times with rate $\lambda$ and 
given exponential service times with rate $\mu$ and the FCFS service policy, the probability of queueing the request (instead of forwarding to a public cloud) is
\begin{equation*} %\label{Eq:forward}
P^{\textit{NF}}(q_i,N_i,Q_i) = \begin{cases}
1 - \sum\limits_{j=0}^{q_i - N_i} \frac{e^{-N_i\mu Q_i}\cdot(N_i\mu Q_i)^j}{j!} & \text{if $q_i \ge N_i$,}\\
1 & \text{if $q_i < N_i$.}
\end{cases}
\end{equation*}%
In particular, $P^{\textit{NF}}(q_i,N_i,Q_i)$ is less than one if the request cannot be served immediately upon arrival (i.e., $q_i \ge N_i$).

At the steady state, the expected probability of forwarding a new request to public clouds is then $P^{F} = \left(\sum_{k=N_i}^{\infty} (1 - P^{\textit{NF}}(k,N_i,Q_i)) \cdot \pi_{k} \right)$, where $\pi_{k}$ is the steady-state probability of having $k$ requests in the system.
Then, the expected rate at which VM requests are forwarded to public clouds is $\overline{P_i^{0}} = \lambda \cdot P^{F}$, which can be used in Eq. (\ref{Eq:cost}) to compute the cost for SCs not sharing resources, i.e., with $\overline{O_i^{0}}=\overline{I_i^{0}}=0$.

\subsection{Detailed Model for SC federation} \label{Sec:initial_model}

%After we can estimate the degenerate case, we start to focus on estimating the performance parameters for the cooperative.
The model of a federation with sharing is complex.
Given a federation of $K$ SCs, each of which will share a maximum of $S_i$ VMs for $i=1,\dots,K$, our goal is to estimate the performance parameters $\overline{P_i^{S_i}}$, $\overline{O_i^{S_i}}$, and $\overline{I_i^{S_i}}$ for each SC~$i$.
To accurately estimate these parameters, we need to consider the interaction among SCs in the federation.
%In order to compute the cost in Eq.~\eqref{Eq:cost} for each SC, we need a performance model to estimate the performance parameters: $\overline{P_i^{S_i}}$, $\overline{O_i^{S_i}}$, and $\overline{I_i^{S_i}}$, based on each SC sharing decision, which is the maximum number of VMs ($S_i$, $i=1,\dots,K$) to share.
%
%A natural model of the $K$ SCs to consider is a continuous-time Markov chain (CTMC), $\mathcal M$, with the following state space $\mathcal S$:
One approach is to build a continuous-time Markov chain (CTMC), $\mathcal M$, with the following state space $\mathcal S$:
\begin{align*}
\mathcal S = & \;\{ (q_1,s_{1,1},\dots,s_{1,K},\;\dots,\;q_K,s_{K,1},\dots,s_{K,K}) \mid q_i\ge 0,\\
& \quad 0\le s_{i,j} \le S_j, \,s_{i,i} = \sum_{j\neq i} s_{j,i} \le S_i, \text{for } i=1,\dots,K \},
\end{align*}%
where $q_i$ is the number of requests from SC~$i$'s customers that are either queued or in service at SC~$i$, $s_{i,i}$ is the number of VMs at SC~$i$ serving requests from other SCs, and $s_{i,j}, i\neq j$ is the number of VMs at SC $j$ being used by SC~$i$.

Transition rates between states of $\mathcal M$ can be assigned so as to implement the probabilistic forwarding mechanism of the model for new arrivals, and service of queued requests.
Table~\ref{Tabl:full_state} reports the transition structure for the detailed model $\mathcal M$ introduced in Section~\ref{Sec:initial_model}. 
Transitions are given for SC $i$ from a generic state
\[(q_1,s_{1,1},...,s_{1,K},...,q_i,s_{i,1},...,s_{i,K},...,q_K,s_{K,1},...,s_{K,K}).\]
The transition rates include $P^F(V_i, n_i, Q_i)$, which is the probability that a request is forwarded to a public cloud when $n_i$ requests are queued at SC~$i$, all of its $V_i\leq N_i$ available VMs are currently busy, and the maximum allowed waiting time by the SLA is $Q_i$ (see Section~\ref{Sec:forward_model} for a detailed definition).
We also assume a load balancing mechanism in the model: SC $i$ determines with which SC~$j$ to share an idle VM by choosing (uniformly at random) among those SCs with the highest number of queued requests.

\begin{table*}
	\centering
	\scalebox{0.75}{
		\begin{tabularx}{24cm}{|>{\centering\arraybackslash}X|>{\centering\arraybackslash}X|>{\centering\arraybackslash}X|}
			\hline
			Next State & Rate & Condition for Transition\\
			\hline
			\hline
			$(q_1,s_{1,1},...,s_{1,K},...,q_i+1,s_{i,1},...,s_{i,K},$\hspace{1cm} & \multirow{2}{\hsize}{\centering$P(q_i, s_{i,i})\lambda_i$} & \multirow{2}{\hsize}{\centering$ (q_i + s_{i,i} < N_i)\vee (q_j + s_{j,j} \ge N_j, \forall j\ne i)$}\\
			\hspace{4cm}$...,q_K,s_{K,1},...,s_{K,K})$ & ~ & ~ \\
			\hline
			$(q_1,s_{1,1},...,s_{1,K},...,$ & \multirow{4}{\hsize}{\centering$\displaystyle\frac{\lambda_i}{|K|}$} & $ (q_i + s_{i,i} \ge N_i)\wedge $\\
			$q_i,s_{i,1},...,s_{i,j}+1,...,s_{i,K},...,$ & ~ & $( L = \{(q_l,s_{l,l})\mid q_l+s_{l,l}<N_l,s_{l,l}<S_l\}, \forall l\ne i) \wedge $\\
			$q_j,s_{j,1},...,s_{j,j}+1,...,s_{j,K},...,$ & ~ & $ (K = \{(q_k,s_{k,k})\mid q_k+s_{k,k}=\min_{L} (q_l+s_{l,l})\})$\\
			$q_K,s_{K,1},...,s_{K,K})$ & ~ & $ \wedge (q_j, s_{j,j}) \in K$\\      
			\hline
			$(q_1,s_{1,1},...,s_{1,K},...,q_i-1,s_{i,1},...,s_{i,K},...,$ & \multirow{2}{\hsize}{\centering$\min((N_i-s_{i,i}), q_i)\mu$} & \multirow{2}{\hsize}{\centering$ (q_i + s_{i,i} > N_i)  \vee (q_j + s_{j,j} \le N_j, \forall j\ne i)$}\\
			$q_j,s_{j,1},...,s_{j,j},...,s_{j,K},...q_K,s_{K,1},...,s_{K,K})$ & ~ & \\
			\hline
			$(q_1,s_{1,1},...,s_{1,K},...,$ & \multirow{4}{\hsize}{\centering$\displaystyle\frac{\min((N_i-s_{i,i}), q_i)\mu}{|K|}$} & $ (q_i + s_{i,i} \le N_i)\wedge (s_{i,i} < S_i) \wedge$\\
			$q_i-1,s_{i,1},...,s_{i,j}+1,...,s_{i,K},...,$ & ~ & $( L = \{(q_l,s_{l,l})\mid q_l+s_{l,l}>N_l\}, \forall l\ne i) \wedge $\\
			$q_j-1,s_{j,1},...,s_{j,j},...,s_{j,K},...,$ & ~ & $ (K = \{(q_k,s_{k,k})\mid q_k+s_{k,k}=\max_{L} (q_l+s_{l,l})\}) $\\
			$q_K,s_{K,1},...,s_{K,K})$ & ~ & $ \wedge (q_j, s_{j,j}) \in K$\\
			\hline
			$(q_1,s_{1,1},...,s_{1,K},...,q_i,s_{i,1},...,s_{i,j}-1,...,s_{i,K},$ & \multirow{2}{\hsize}{\centering$s_{i,j}\mu$} & \multirow{2}{\hsize}{\centering$(q_j + s_{j,j} > N_i) \vee (q_k + s_{k,k} \le N_k, \forall k \ne j)$}\\
			$...,q_j,s_{j,1},...,s_{j,j}-1,...,s_{j,K},...q_K,s_{K,1},...,s_{K,K})$ & ~ & \\
			\hline
			$(q_1,s_{1,1},...,s_{1,K},...,q_i,s_{i,1},...,s_{i,j}-1,...,s_{i,K},...,$ & \multirow{4}{\hsize}{\centering$\displaystyle\frac{s_{i,j}\mu}{|K|}$} & $ (q_j + s_{j,j} \le N_j)\wedge$\\
			$q_j,s_{j,1},...,s_{j,j},...,s_{j,K},...,$ & ~ & $( L = \{(q_l,s_{l,l})\mid q_l+s_{l,l}>N_l\}, \forall l\ne i) \wedge $\\
			$q_m,s_{m,1},...,s_{m,j}+1,...,s_{m,K},...,$ & ~ & $ (K = \{(q_k,s_{k,k})\mid q_k+s_{k,k}=\max_{L} (q_l+s_{l,l})\}) $\\
			$q_K,s_{K,1},...,s_{K,K})$ & ~ & $ \wedge (q_m, s_{m,m}) \in K$\\
			\hline
		\end{tabularx}
	}\vspace{.5cm}
	\caption{State transitions in $\mathcal M$ from state $(q_1,s_{1,1},...,s_{1,K},...,q_i,s_{i,1},...,s_{i,K},...,q_K,s_{K,1},...,s_{K,K})$}
	\label{Tabl:full_state}
\end{table*}

Although solving $\mathcal M$ could give us an accurate prediction of all performance characteristics required in Eq.~\eqref{Eq:cost}, the corresponding state space $\mathcal S$ grows exponentially with $K$.
Since re-computation of sharing decisions is needed when significant changes in workload or resource availability occur, a model with a more efficient solution is desirable.
%Thus, we propose an approximate (yet accurate) model in Section~\ref{Sec:performance_model} that is able to produce results in (near) real-time.
%
Moreover, solving for $\mathcal M$ requires obtaining detailed SC information (such as the arrival rate, the number of overall VMs, and the SLA) that SCs might not want to release.
Thus, each SC should be able to compute the model in a decentralized manner and release as little information as possible.

\subsection{Approximate Model for SC Federation}\label{Sec:approximate_model}

%\begin{figure}
%	\centering
%	\includegraphics[width=0.40\textwidth]{new_cloudshare_model} 
%	\caption{An abstract of our approximate model for $K$ SCs}
%	\label{Fig:approximate_model}
%\end{figure}

In this section, we focus on an approximate model that can be solved quickly (as system conditions, such as workload, change) and in a decentralized manner  (without releasing too much information to other SCs), but also yields sufficiently accurate results, in order to produce appropriate sharing decisions.
By analyzing the detailed model $\mathcal M$, we realize that using $\mathcal M$ allows estimation of performance parameters for all SCs in the federation simultaneously;
however, in realistic scenarios, each SC computes its own performance parameters to estimate its cost assuming that other SCs' sharing decisions are fixed; thus, there is no need for the performance model to simultaneously output results for all SCs.
Moreover, since we assume that the same cost is charged by all SCs for shared VMs,
%our cost estimation does not consider different costs for using VMs from individual SCs, 
an SC does not need to distinguish the source or destination of shared VMs.
Therefore, we propose a hierarchical approximate model that computes performance parameters iteratively.

%We use a $K$ SCs environment as an example to describe our approximate model.
Given a federation of $K$~SCs, we consider each SC~$i=1,\dots,K$ in sequence, where SC $K$ is the SC of interest, which we refer to as \emph{target SC} in the rest of the paper. At each step, we build and analyze a Markovian model $\mathcal M^i$ where only SCs~$\{1,\dots,i\}$ can access shared resources of the federation. The model $\mathcal M^i$ takes into account the solution of $\mathcal M^{i-1}$ and refines it to include also SC~$i$.
For example, in model $\mathcal M^1$, the first SC has exclusive access to all shared resources of the federation; in $\mathcal M^2$, only SC~$1$ and SC~$2$ utilize shared resources from all SCs, but VM allocations in $\mathcal M^1$ are taken into account.
We repeat this process until reaching the target SC.
In this approach, since SC~$i$ only needs the solution of $\mathcal M^{i-1}$ to build $\mathcal M^i$, we allow SCs not to leak sensitive information on capacity and SLAs.
In the following, we give a detailed description of $\mathcal M^{i},\, 1\le i\le K$ and of its solution.

\smallskip\noindent\textbf{State Space $\mathcal S_i$ for $\mathcal M^{i}$}.
%Depending on which level of recursiveness, $\mathcal M^i$ would be constructed with different state spaces. If there are total $K$ SCs, the state space $\mathcal S^i$ for $\mathcal M^i$ is
%In our approximate model, the state space $\mathcal S^i$ for $\mathcal M^i$ 
The state space $\mathcal S_i$ of $\mathcal M^{i}$ is
{\small
\begin{align*}
\mathcal S^i = \{ (q_i,s_i,o_i,a_i) \mid q_i\geq 0,\, 0 \leq s_i\le S_i,\, 0 \le o_i+a_i \le B_i\},
\end{align*}}%
where $q_i$ is the total number of requests at SC~$i$ (queued or in service), $s_i$ is the number of VMs of SC~$i$ currently used to serve requests from SCs~$\{1,\dots,i-1\}$, $o_i$ is the number of VMs from other SCs currently used by SC~$i$, and $a_i$ is the number of shared VMs used by SCs in $\mathcal M^{i-1}$.
%
%Here, we use $s_i$ and $o_i$ to represent the number of shared VMs shared by and used by SC~$i$. 
%
%Since our recursive model $\mathcal M^i$ can only focus on the state of $\mathcal S^{i+1}$, $a_i$ and $w_i$ are used to explore the use of VMs from bottom level recursiveness.
%
%(However, in $\mathcal S^K$, $a_i = w_i = 0$ because there is no lower model for SC~$K$.)
%
Given that there are at most $N_i$ VMs in SC~$i$, $\max(0,q_i-(N_i-s_i))$ requests are waiting at SC~$i$; moreover, $s_i$ is bounded by $S_i$, the maximum number of VMs shared by SC~$i$.
Since $\mathcal M^{i}$ includes SCs~$\{1,\dots,i\}$ and SC~$i$ is the target SC in $\mathcal M^{i}$, we use $o_i$ to record the number of shared VMs (not from SC~$i$) used by SC~$i$, and we use $a_i$ to record the number of shared VMs (not from SC~$i$) used by SCs~$\{1,\dots,i-1\}$; thus, $o_i + a_i$ is bounded by $B_i = \sum_{j\ne i} S_j$, the maximum number of VMs shared by SCs~$\{1,\dots,K-1\}$.
\smallskip\noindent\textbf{State Transitions}.
%\begin{figure}
%	\centering
%	\includegraphics[width=0.4\textwidth]{state_trans} 
%	\caption{An abstract of finding the next state}
%	\label{Fig:state_trans}
%\end{figure}
%
%As illustrated in Fig. \ref{Fig:approximate_model}, %$\mathcal M^{i}$ deploy its VM allocations based on the VM allocations in $\mathcal M^{i-1}$.
%we use the information of VM allocations in $\mathcal M^{i-1}$ to build $\mathcal M^{i}$.
%
VM allocations in $\mathcal M^{i-1}$ affect the results of new states in $\mathcal M^{i}$ after state transitions.
Each state transition happens in the period of time between two events (referred to as \emph{inter-event period} in the rest of paper), each of which can be a request arrival or a service completion instance.
%
%As illustrated in Fig. \ref{Fig:state_trans}, 
During an inter-event period, each state in $\mathcal M^{i}$ can increase the number of VMs shared by SC~$i$ due to SCs in $\mathcal M^{i-1}$ allocating VMs in SC~$i$; similarly, the number of requests queued at SC~$i$ can decrease due to service completions in $\mathcal M^{i-1}$, which allow SC~$i$ to utilize shared VMs.
%
%interactions between models $\mathcal M^i$ and $\mathcal M^{i+1}$ correspond to requests for idle VMs: after request arrival instances at SC~$i$, the number of VMs shared by SC~$i$ can increase due to VM requests from all SC $j, \forall j>i$ (thereby reducing the number of VMs available at $i$); similarly, the number of requests queued at SC~$i$ can decrease due to service completions at all lower level SCs, SC $j, \forall j>i$, which allow SC~$i$ to forward some of its VM requests to all lower level SCs.
%
%SC~$i$ updates its state at request arrival or service completion instances; the new state for SC~$i$ must also reflect the expected state change due to VMs allocated at SC~$i$ by lower-level SCs, and by SC~$i$ at lower-level SCs during the previous inter-event period.
%
Thus, the probability of going to any destination state from a state of $\mathcal M^{i}$ depends on the probability of being at a specific state in $\mathcal M^{i-1}$. %, which represent the VM allocations in other $i-1$ SCs.
Here, we define three \emph{interaction probability vectors} representing the probability of moving from each state $(q_{i}, s_{i}, o_{i}, a_{i})$ of $\mathcal M^{i}$ to any other state of $\mathcal M^{i}$ when an event happens, based on the interaction probabilities computed for $\mathcal M^{i-1}$:
\begin{itemize}
	\item $P^A(q_{i}, s_{i}, o_{i}, a_{i})$ for an inter-event period preceding an arrival instance;
	
	\item $P^D_{loc}(q_{i}, s_{i}, o_{i}, a_{i})$ for an inter-event period preceding a local departure instance;
	
	\item $P^D_{rem}(q_{i}, s_{i}, o_{i}, a_{i})$ for an inter-event period preceding the remote departure instance of a VM allocated at other SCs by SC~$i$.

\end{itemize}
The detailed computation of these interaction probability vectors is described below.

Let $a_{loc}$ represent the number of VMs shared by SC~$i$ and allocated by SCs~$\{1,\dots,i-1\}$ in $\mathcal M^{i-1}$, and let $a_{rem}$ represent the number of VMs shared by all other SCs (except SC~$i$) and allocated by SCs~$\{1,\dots,i-1\}$ in $\mathcal M^{i-1}$, respectively.
Then, given a state in $\mathcal M^{i-1}$, which can produce the pair $(a_{loc}, a_{rem})$, $P^A(q_{i}, s_{i}, o_{i}, a_{i})_{(a_{loc}, a_{rem})}$, $P^D_{loc}(q_{i}, s_{i}, o_{i}, a_{i})_{(a_{loc}, a_{rem})}$, and $P^D_{rem}(q_{i}, s_{i}, o_{i}, a_{i})_{(a_{loc}, a_{rem})}$ represent the probability of allocating VMs $(a_{loc}, a_{rem})$ in vectors $P^A(q_{i}, s_{i}, o_{i}, a_{i})$, $P^D_{loc}(q_{i}, s_{i}, o_{i}, a_{i})$, and $P^D_{rem}(q_{i}, s_{i}, o_{i}, a_{i})$, respectively, after an event in the state $(q_{i}, s_{i}, o_{i}, a_{i})$ of $\mathcal M^{i}$.
%
%represents the probability of allocating ($a_{loc}$, $a_{rem}$) VMs in $\mathcal M^{i-1}$ when the state in $\mathcal M^{i}$ is $(q_{i}, s_{i}, o_{i}, a_{i})$ and an arrival happens to SC~$i$.
%
The legal combinations of the pairs $(a_{loc},a_{rem})$ are determined by the current state $(q_{i}, s_{i}, o_{i}, a_{i})$ of $\mathcal M^{i}$, as described below. 
For simplicity, in the rest of paper we use ${P^A}_{(a_{loc}, a_{rem})}$, ${P^D_{loc}}_{(a_{loc}, a_{rem})}$, and ${P^D_{rem}}_{(a_{loc}, a_{rem})}$ to represent the probability of VM allocations in $\mathcal M^{i-1}$ for each state $(a_{loc},a_{rem})$, given the state $(q_{i}, s_{i}, o_{i}, a_{i})$ of $\mathcal M^{i}$.
%to represent the probability of allocating VMs ($a_{loc}$, $a_{loc}$) in vectors $P^A(q_{i}, s_{i}, o_{i}, a_{i})$, $P^D_{loc}(q_{i}, s_{i}, o_{i}, a_{i})$, and $P^D_{rem}(q_{i}, s_{i}, o_{i}, a_{i})$ respectively, after an event happens to the state $(q_{i}, s_{i}, o_{i}, a_{i})$ in $\mathcal M^{i}$.
%We enumerate all possible combinations of $a$ and $b$ for all transitions in order to include all cases.
%
%Moreover, in order to meet its SLAs, SC~$i$ has a probability $P^{F}(q_1,V_1,Q_1)$ of forwarding requests to a public cloud if they cannot be served before the deadline $Q_i$ (see Section \ref{Sec:forward_model}). 

%We start to give a detailed description and formal structure for the transitions of $\mathcal M^i$.
%
%The state space $\mathcal S^i$ for $\mathcal M^i$ follows the bounds defined in Section \ref{Sec:markov-model}.
%
%Since, $\mathcal M^{K}$ has no lower level model to be considered for its transitions, which is illustrated in Fig. \ref{Fig:state_trans}, in the following transition cases, we will discuss $\mathcal M^{K}$ and $\mathcal M^{i},\,1\le i<K$ separately.

\smallskip\noindent\textbf{Transitions for $\mathcal M^{1}$}. In $\mathcal M^{1}$, there is only one SC, and no other model affecting the transitions; thus, $s_1 = a_1 = 0$, and
\footnotesize
\begin{align*}
&(q_1, 0, o_1, 0) \xrightarrow{\lambda} (q_1+1, 0, o_1, 0) && \text{if $q_1 < {N}_1$}\\
&(q_1, 0, o_1, 0) \xrightarrow{\lambda} (q_1, 0, o_1+1, 0) && \text{if $q_1 \ge {N}_i \wedge o_1 < {B}_1$}\\
&(q_1, 0, o_1, 0) \xrightarrow{\lambda\cdot P^{\textit{NF}}(q_1,N_1,Q_1)} (q_1+1, 0, o_1, 0) && \text{if $q_1 \ge {N}_i \wedge o_1 = {B}_1$}\\
&(q_1, 0, o_1, 0) \xrightarrow{\min(q_1, N_1)\cdot\mu} (q_1-1, 0, o_1, 0) && \text{if $q_1 > 0$}\\
&(q_1, 0, o_1, 0) \xrightarrow{o_1\cdot\mu} (q_1, 0, o_1-1, 0) && \text{if $o_1 > 0$}
\end{align*}
\normalsize

\smallskip\noindent\textbf{Transitions for $\mathcal M^{i}$}. 
Any transition in $\mathcal M^{i}$ with $i>1$ depends on interaction probability vectors for $\mathcal M^{i-1}$.
%
%Here, for simplicity, we use ${P^A}_{(*)}$, ${P^D_{loc}}_{(*)}$, and ${P^D_{rem}}_{(*)}$ to represent the probability of being at state $(q_{i-1}, s_{i-1}, o_{i-1}, a_{i-1})$ in vectors $P^A(q_{i}, s_{i}, o_{i}, a_{i})$, $P^D_{loc}(q_{i}, s_{i}, o_{i}, a_{i})$, and $P^D_{rem}(q_{i}, s_{i}, o_{i}, a_{i})$ respectively, after an event happens to the state $(q_{i}, s_{i}, o_{i}, a_{i})$ in $\mathcal M^{i}$.
%
Given any pair $(a_{loc}, a_{rem})$ from states in $\mathcal M^{i-1}$, the transitions corresponding to a request arrival instance at state $(q_{i}, s_{i}, o_{i}, a_{i})$ in $\mathcal M^{i}$ fall into one of the following cases:

\smallskip\noindent $C_1$: 
The new request can use a VM at SC~$i$ when there is at least one free VM at SC~$i$, even after considering $a_{loc}$ and $a_{rem}$ from $\mathcal M^{i-1}$ during the arrival period: %Given the current state $(q_i, s_i, o_i, a_i)$, 
%The formal structure for this transition is
\begin{align*}
(q_i, s_i, o_i, a_i) \xrightarrow{\lambda \cdot {P^A}_{(a_{loc},a_{rem})}} (q_i+1, a_{loc}, o_i, a_{rem})
\end{align*}
for all $q_i + a_{loc} < {N}_i$ such that $(a_{loc} \le {S}_i) \wedge (o_i + a_{rem} \le {B}_i)$. 
% We will discuss $a$ and $b$, which are the VM allocations from $i-1$ SCs at state $(q_{i-1}, s_{i-1}, o_{i-1}, a_{i-1})$ in $\mathcal M^{i-1}$, later.

\smallskip\noindent $C_2$:
The new request uses a VM from other SCs. This situation arises when SC~$i$ has no idle VMs prior to this arrival instance, but other SCs can provide at least one VM during the preceding inter-event period:
\begin{align*}
(q_i, s_i, o_i, a_i) \xrightarrow{\lambda \cdot {P^A}_{(a_{loc},a_{rem})}} (q_i, a_{loc}, o_i+1, a_{rem})
\end{align*}
for all $q_i + a_{loc} \ge {N}_i$ and $o_i+a_{rem}+1 \le {B}_i$.

\smallskip\noindent $C_3$:
The new request must be queued or forwarded to a public cloud due to no available shared VMs in the federation,
%. This case corresponds to the situation in which 
where all VMs have been occupied during the previous or current inter-event period by requests from other SCs: % The formal structure for this transition is 
\begin{align*}
(q_i, s_i, o_i, a_i) \xrightarrow{\substack{\lambda \cdot {P^A}_{(a_{loc},a_{rem})} \\ \cdot P^{\textit{NF}}(q_i,V_i,Q_i)}} (q_i+1, a_{loc}, o_i, a_{rem})
\end{align*}
for all $q_i + a_{loc} \ge {N}_i$ and $o_i+a_{rem}={B}_i$. $V_i = N_i - s_i + o_i$ is the number of VMs in the federation currently used by SC~$i$.

Given any pair $(a_{loc}, a_{rem})$ for states in $\mathcal M^{i-1}$, the transitions corresponding to a service completion instance at SC~$i$ for its own customers fall into one of the following cases:

%\wedge (\exists c, o_i - c \ge 0) \wedge ((o_i - c)+(o_{i-1}-) ) \wedge (a + b = min(q_{i-1}+s_{i-1}, {S}_{i-1})+o_{i-1}+a_{i-1})$.
%
%Here, ${N'}_{i} = N_{i}-S_{i}+{S'}_{i}$ represents the available VMs at SC~$i$ after sharing VMs to higher level SCs (SC $1$...SC $i-1$), and $min(q_{i+1}+s_{i+1}, {S'}_{i+1})$ represents the number of shared VMs from SC $i+1$ used by SC $i+1$$...$SC $K$.
%
%Thus, the value of $k_1 + k_2$ equals to the overall number of shared VMs not used by SC~$i$ but used by SC $i+1$...SC $K$.
%
%The value of ${w'}_i$ will be set to $1$ if there is any requests queued in lower level SCs:
%\[
%{w'}_i =\begin{cases}
%1 & \text{if $(w_{i+1}=1) \vee (q_{i+1} - ({N'}_{i+1}-s_{i+1})>0)$;}\\
%0 & \text{otherwise,}
%\end{cases}
%\]
%where ${N'}_{i+1} = N_{i+1}-S_{i+1}+{S'}_{i+1}$.
%
%For $\mathcal M^K$, since there is not lower lever SCs, $s_K = a_K = w_K = 0$.

\smallskip\noindent $C_4$:
The departure is from VMs of SC~$i$ used by SC~$i$ itself. 
If there is at least one job queued in SC~$i$, the freed VM will be used by SC~$i$ directly:
\begin{align*}
	(q_i, s_i, o_i, a_i) \xrightarrow{L_i\cdot \mu \cdot {P^D_{loc}}_{(a_{loc},a_{rem})}} (q_i-1, a_{loc}, o_i,a_{rem}),
\end{align*}
where $L_i = \min(q_i, {N}_i-s_i)$ is the number of VMs from SC~$i$ used by SC~$i$, for all $q_i+a_{loc} > {N}_i$. However, if there are no queued requests in SC~$i$, the freed VM will be assigned to other SCs with queued jobs:
%. The formal structure for this transition is
\begin{align*}
(q_i, s_i, o_i, a_i) \xrightarrow{L_i\cdot \mu \cdot {P^D_{loc}}_{(a_{loc},a_{rem})}} (q_i-1, a_{loc}+1, o_i,a_{rem})
\end{align*}
for all $q_i + a_{loc} \le {N}_i$.
% if the state in $\mathcal M^{i-1}$ has queued requests. 
If other SCs do not have queued requests, the transition has the same form as in the previous case for queued requests at SC~$i$.

\smallskip\noindent $C_5$:
The departure is from VMs of other SCs allocating to SC~$i$. If there are no queued jobs in any SCs, the freed VM will be returned directly: 
\begin{align*}
(q_i, s_i, o_i, a_i) \xrightarrow{o_i \cdot \mu \cdot {P^D_{rem}}_{(a_{loc},a_{rem})}} (q_i, a_{loc}, o_i-1, a_{rem})
\end{align*}
for all $q_i + a_{loc} \le {N}_i$. If at least one request is queued in SCs~$\{1,\dots,i-1\}$, SC~$i$ must share the VM:%the formal structure for the transition is
\begin{align*}
(q_i, s_i, o_i, a_i) \xrightarrow{o_i \cdot \mu \cdot {P^D_{rem}}_{(a_{loc},a_{rem})}} (q_i, a_{loc}, o_i-1, a_{rem}+1).
\end{align*}
However, if the above conditions are not satisfied and there is at least one job queued in SC~$i$, the VM will still be assigned to SC~$i$, for all $q_i+a_{loc} > {N}_{i}$: %. The formal structure for this transition is
\begin{align*}
(q_i, s_i, o_i, a_i) \xrightarrow{o_i \cdot \mu \cdot {P^D_{rem}}_{(a_{loc},a_{rem})}} (q_i-1, a_{loc}, o_i, a_{rem}).
\end{align*}

%The values of $a$ and $b$ are computed according to specific state $(q_{i}, s_{i}, o_{i}, a_{i})$ in $\mathcal M^{i}$ and state $(q_{i-1}, s_{i-1}, o_{i-1}, a_{i-1})$ in $\mathcal M^{i-1}$, and satisfy the following conditions:
%\begin{enumerate}
%	\item $a$ is the number of shared VMs in SC~$i$ used by SCs in $\mathcal M^{i-1}$, thus the value of $a$ is the summation of the subset of $s_{i-1}$ and the subset of $a_{i-1}$.
%	\item $b$ is the number of shared VMs not from SC~$i$ currently used by SCs in $\mathcal M^{i-1}$, thus the value of $b$ is the summation of all occupied VMs except ones used from SC~$i$ (the value of $a$).
%\end{enumerate}
%We enumerate all possible combinations of $a$ and $b$ for all transitions in order to include all cases.

%\subsubsection{Interaction Probabilities} \label{Sec:interation_prob}
\smallskip\noindent\textbf{Interaction Probabilities}.
As mentioned above, the interaction probabilities describe the probability of different VM allocations from SCs in $\mathcal M^{i-1}$ during an inter-event period of~$\mathcal M^{i}$.
%
%\sunghan{Steady-state and transient analysis (using \emph{Uniformization} \cite{St95} to transform the CTMC into a discrete-time Markov chain of the total number of occupied VMs in $\mathcal M^{i-1}$ can be leveraged for the computation of interaction probabilities between $\mathcal M^{i}$ and $\mathcal M^{i-1}$.}
%
To compute transient probabilities, which describe transient changes in the number of VM allocations in CTMC $\mathcal M^{i-1}$ over inter-event periods at SC~$i$, 
%We use steady-state probabilities as an initial state distribution to 
%
%To , 
we use the method of \emph{uniformization} \cite{St95} to transform the CTMC into a discrete-time Markov chain and a Poisson process as follows: 
given the infinitesimal generator $Q^{i-1}$,
\begin{itemize}
	\item the rate of the Poisson process is
	$\gamma \geq \max_j \big|q_{jj}^{i-1}\big|$,
	\item the transition matrix of the DTMC is
	$P^{i-1} = I + \frac{1}{\gamma} Q^{i-1}$.
\end{itemize}
Then, the transient probability vector $p^{i-1}(t)$ for $\mathcal M^{i-1}$ can be computed for all $t \geq 0$ as $p^{i-1}(t) = p_0 P^{i-1}(t)$, 
where $P^{i-1}(t) = \sum_{k=0}^{\infty} \frac{e^{-\gamma t}(\gamma t)^k}{k!} (P^{i-1})^k$
is the matrix of transition probabilities for the CTMC (for a given precision $\epsilon$, the summation can be truncated using the Fox and Glynn method \cite{fox88}).
By letting $p_0$ be equal to the initial state distribution at any time instance, we can compute transient state changes of $\mathcal M^{i-1}$.

\begin{figure}
	\centering
	\includegraphics[width=0.4\textwidth]{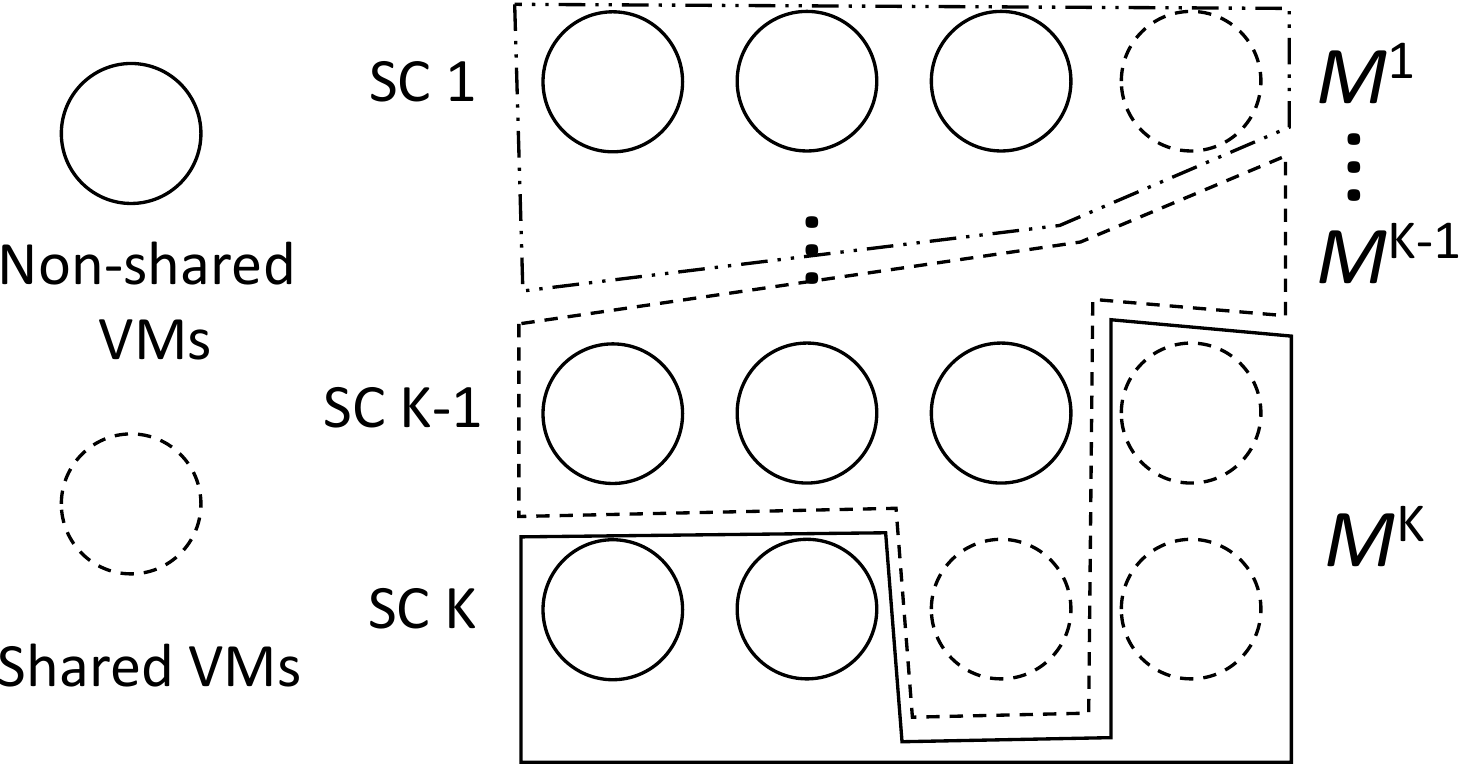} % previous_reference
		\caption{Example of allocation constraints for a state $(q_{i}, s_{i}, o_{i}, a_{i})$ in $\mathcal M^{i}$}
	\vspace{-1em}
	\label{Fig:vm_allocation}
\end{figure}

\smallskip\noindent\textbf{Initial State Distribution}.
%Using transient analysis helps us understand the probability of being at each state after a specific duration from an initial state distribution ($p_0$).
%
%However, when we compute the interaction probability vectors for $\mathcal M^{i}$, 
The initial distribution of $\mathcal M^{i-1}$ depends on VM allocations in the current state of $\mathcal M^{i}$.
For instance, as illustrated in Fig. \ref{Fig:vm_allocation}, when $S_i=3$ and SC~$i$ uses $2$ of its shared VMs in state $(q_{i}, s_{i}, o_{i}, a_{i})$ of $\mathcal M^{i}$, only up to $1$ of the remaining SC~$i$'s shared VMs can be allocated to others in $\mathcal M^{i-1}$.
%
%the probability of each state in $\mathcal S^{i+1}$ is restricted by $\mathcal M^{i}$.
%
%Thus, the starting point for $\mathcal M^{i+1}$ to use transient analysis depends on the current VM occupations in state of $\mathcal M^{i}$.
%
%For instance, as illustrated in Fig. \ref{Fig:previous_reference}, if the state in $\mathcal M^{K-1}$ has one free VM left in SC~$K-1$, the starting point of $\mathcal M^{K}$ should not occupied this free VM in SC~$K-1$; otherwise, according to the definition in Section \ref{Sec:state_trans}, the state in $\mathcal M^{K-1}$ should mark this VM used (since $\mathcal S^{K-1}$ will record the VM allocations in the lower-level model, $\mathcal M^{K}$).
%

We represent the initial distribution of $\mathcal M^{i-1}$ over its state space $\mathcal S^{i-1}$ as $\pi^X_{[(q_i, s_i, o_i, a_i)]}$, where $[(q_i, s_i, o_i, a_i)]$ is the subset of states in $\mathcal S^{i-1}$ that satisfy VM allocation constraints for a given state $(q_i, s_i, o_i, a_i)$ of $\mathcal M^{i}$.
The initial state distribution $\pi^X_{[(q_i, s_i, o_i, a_i)]}$ of $\mathcal M^{i-1}$ is computed from its steady-state probabilities by considering only states $[(q_i, s_i, o_i, a_i)]$ and renormalizing their probability masses. 
Then, interaction probability vectors for $\mathcal M^{i-1}$ and $\mathcal M^{i}$ are given by the product of the initial state distribution and transient state change during the average inter-arrival time or departure time:
%
%\begingroup\makeatletter\def\f@size{8}\check@mathfonts
%\def\maketag@@@#1{\hbox{\m@th\large\normalfont#1}}%
\begin{align*}
P^A(q_{i}, s_{i}, o_{i}, a_{i})&= 
\Big(\pi^X_{[(q_i, s_i, o_i,a_i)]}P({\scriptstyle\frac{1}{\lambda_i}})\Big) \\
P^D_{loc}(q_{i}, s_{i}, o_{i}, a_{i})&= 
\Big(\pi^X_{[(q_i, s_i, o_i,a_i)]}P({\scriptstyle\frac{1}{L_i \mu}})\Big) \\
P^D_{rem}(q_{i}, s_{i}, o_{i}, a_{i})&= 
\Big(\pi^X_{[(q_i, s_i, o_i,a_i)]}P({\scriptstyle\frac{1}{o_i \mu}})\Big)
\end{align*} %\endgroup
where $L_i$ is the number of local busy VMs in $(q_i, s_i, o_i, a_i)$.
The initial state distribution $\pi^X_{[(q_i, s_i, o_i, a_i)]}$ is computed through the concept of Conditional Probability Distribution \cite{billingsley2008probability}. % (see Appendix \ref{Sec:conditional_probability}).
%given $X$, a discrete random variable over $\mathbb{N}=\{0,1, \dots\}$, and $\pi^X$, its probability mass function, the conditional probability distribution for subset $Y\subseteq X$ is:
%\begin{align}
%\pi^X(k) = \begin{cases}
% P_X(k | k\in Y) = \frac{P(X=k \wedge k\in Y)}{P(k\in Y)} & \text{if $ k\in Y$,}\\
% 0 & \text{otherwise.}
%\end{cases}
%\end{align}

%Truncated Distribution \cite{johnson1970discrete}: given $X$, a discrete random variable over $\mathbb{N}=\{0,1, \dots\}$, and $\pi^X$, its probability mass function, the truncated probability for subset $Y\subseteq X$ is:
%\begin{align}\label{eq:truncation}
%\pi^X_Y(k) = \begin{cases}
%\frac{\pi^X(k)}{\sum_{k\in Y}\pi^X(i)} & \text{if $ k\in Y$,} \\
%	0 & \text{otherwise.}
%\end{cases}
%\end{align}

%To obtain the initial state distribution for possible states, we use the following definitions to restrict the state space of a DTMC from $\mathbb{N}=\{0,1, \dots\}$ to a finite interval $[x_1, x_2] \subset \mathbb{N}$.
%
%\begin{definition}[Definition (Truncated random variable).]
%Let $X$ be a discrete random variable over
%$\mathbb{N}=\{0,1, \dots\}$ and let $\pi^X$ be its probability mass
%function. Then, for each $x_1,x_2\in\mathbb{N}$ such that
%$x_1\leq x_2$, the random variable
%\[
%X_{x_1,x_2} := \begin{cases}
%X & \text{if $x_1 \le X \le x_2$,} \\
%0 & \text{otherwise}
%\end{cases}
%\]
%has probability mass function $\pi^X_{x_1,x_2}$ given by
%\end{definition}

\smallskip\noindent\textbf{Performance Parameters}.
Given that $\pi^{i}$ represents the steady-state probabilities of $\mathcal M^{i}$, the performance parameters can be computed as follows:
{\small
\begin{align*}
&\overline{I_i^{S_i}} = \sum s_i \cdot \pi^{i}_{(q_i, s_i, o_i,a_i)},\qquad \overline{O_i^{S_i}} = \sum o_i \cdot \pi^{i}_{(q_i, s_i, o_i,a_i)},\\
&\overline{P_i^{S_i}} = \lambda_i \cdot \left(\sum (1 - P^{\textit{NF}}(q'_i,V_i,Q_i)) \cdot \pi^{i}_{(q_i, s_i, o_i,a_i)} \right),
\end{align*}}
where $q'_i = q_i - (N_i - s_i)$ and $V_i = N_i-s_i+o_i$.

%\subsubsection{Speedup the Model Computation}

%If we follow the procedure illustrated in Fig. \ref{Fig:recur_model}, we have to go through models from $\mathcal M^{2}$ to $\mathcal M^{K}$ for each state of $\mathcal M^{1}$, which has significant time complexity.
%
%The beauty of our recursive model is enabling us to construct models in a bottom-up manner, which starts from constructing $\mathcal M^{K}$ first.
%
%We enumerate all possible cases in $\mathcal M^{K}$ and use them to construct $\mathcal M^{K-1}$.
%
%After constructing all possible cases in $\mathcal M^{K-1}$, we can free all spaces allocated to $\mathcal M^{K}$ and only use $\mathcal M^{K-1}$ to construct $\mathcal M^{K-2}$.
%
%We repeat above steps until we have solve the steady-state probabilities of $\mathcal M^{1}$.

\section{Market-based Model} \label{Sec:market_model}

Next, we develop the empirical market-based model for \emph{SC-Share} to determine appropriate sharing decisions for each SC.
%
%The market-based model is needed to determine appropriate sharing decisions for each SC and uses the performance characteristics computed by the performance model.
%
%From the perspective of an individual SC, it is reasonable that the price for VM usage is an important parameter in considering a sharing policy, where higher VM prices could motivate low utilization SCs to share their resources.
%
%Here, we implement a game to determine proper VM prices so as to provide sufficient incentives for SCs to participate in the cooperative.
%
We first formulate SC utility functions that take performance characteristics (as computed above) into consideration. We then focus on the details of the game and on the notion of market efficiency.

\subsection{SC Utilities} \label{Sec:utility}

%Participating in the cooperative enables an SC to obtain resources and satisfy SLAs at prices cheaper than public clouds, and sell idle resources to other SCs for profit, e.g., the way Amazon AWS sells Spot Instances \cite{awsspot}.
As discussed before, SCs participate in the federation in order to obtain resources and satisfy SLAs at prices cheaper than public clouds, and sell idle resources to other SCs for profit, 
%e.g., the way Amazon AWS sells Spot Instances \cite{awsspot}
similarly to spot instances sold by Amazon AWS \cite{awsspot}. %Thus, an SC will try to minimize its cost for satisfying SLAs.
To this end, we define SC $i$'s utility $U^{S_i}_i$ (see Eq. (\ref{Eq:utility}) below) from the ratio between (a) the change in net cost of an SC when it participates in the federation versus when it does not, and (b) the change in utilization of an SC when it participates in the federation versus when it does not:
%
%Mathematically, we define the utility of SC $i$, , when sharing maximum $S_i$ VMs, as:
%
\begin{equation} \label{Eq:utility}
U^{S_i}_i = \frac{{(\max(C_i^{0} - C_i^{S_i},0))}^2}{{(\rho_i^{S_i} - \rho_i^{0})}^{\gamma}}\qquad 0 \le \gamma \le 1,
\end{equation}
where $C_i^{0}$ is the cost for SC $i$ when it does not participate in the federation, $C_i^{S_i}$ is the cost for SC $i$ when it shares a maximum of $S_i$ VMs, $\rho_i^{0}$ is the system utilization (i.e., the fraction of time that SC~$i$'s VMs are busy) when not participating in the federation, and $\rho_i^{S_i}$ is the utilization of SC $i$ when it shares a maximum of $S_i$ VMs.
It is evident that an SC will try to minimize its cost for satisfying SLAs; thus, we consider the cost reduction as the numerator of Eq. (\ref{Eq:utility}).
We consider the increment in SCs' utilizations (the denominator of Eq. (\ref{Eq:utility})) because SCs always want to keep utilizing their resources in a certain level (the system utilization of SCs should always increase since all of them have to share resources with others in order to participate in the federation).
For instance, an SC would want to increase the amount of shared VMs (i.e., increase its system utilization) to obtain higher profit from the cooperation, but would like to decrease the amount of shared VMs whenever its high system utilization makes it forward more requests to a public cloud (i.e., the rate of cost reduction starts to decrease).
%
%The scenario where an SC is willing to increase its cost but reduce its power consumption through lowering system utilizations will not be addressed here, and is the part of future work.
%
The parameter $\gamma$ in Eq. (\ref{Eq:utility}) reflects the importance SC~$i$ places on utilization, where $\gamma = 0$ means SC~$i$ only considers cost reduction, referred to as ${UF}_0$ in the rest of the paper, and $\gamma = 1$ means SC $i$ considers the marginal cost reduction for utilization changes as the most important factor, referred to as ${UF}_1$ in the rest of the paper ($\gamma = 1$ gives highest importance to utilization increase since $0 < \rho_i^{S_i} - \rho_i^{0} \le 1$).
%
%In this utility function, $\gamma = 1$ means an SC cares the marginal revenue from increasing the system utilization, while $\gamma = 0$ means an SC only focuses on reducing its cost.
%
We choose such structure for $U_i^{S_i}$ so that an SC will always try to reduce cost, and the marginal utility is linear in $(C_i^{0} - C_i^{S_i})$.
In the experiments, we assume that all SCs in the federation have the same value for the $\gamma$ parameter, as different values would produce different scales of utility.

\subsection{Non-Cooperative Game Among SCs} \label{Sec:non_cooperative}

%We first introduce the game setting, and then show (through a simulation-based study) how the system behaves at equilibrium points under computed sharing decisions.

\begin{algorithm}[t]
	\caption{Proposed repeated game among SCs}\label{Algo:noncooperate}
	\textbf{Input:} $C_i^P, C_i^G$, SC $\{1,...,K\}$  \\
	\textbf{Output:} $\{S_1,...,S_K\}$ \\
	%There are $1,...,K$ SCs in the cooperative\;
	SC $i$ has $N_i$ VMs, arrival rate $\lambda_i$ and SLA requirement $Q_i$\;
	In round $r = 0$, SC VM sharing vector is $\{S_1^{(0)},...,S_K^{(0)}\}$\;
	\Do{$\exists i \in \{1,...,K\}, \, S_i^{(r)} \ne S_i^{(r-1)}$}{
		$r = r + 1$\;
		\ForEach{$i \in 1,...,K$} {
			$S_i^{(r)}\leftarrow$ the shared VM number which maximizes \\
			\qquad SC $i$'s utility based on $S_j^{(r-1)},\, \forall j\ne i$, $C_i^P, C_i^G$\;
		}
	}
	$\{S_1^{(r)},...,S_K^{(r)}\}$ is the equilibrium point\;
\end{algorithm}

\smallskip\noindent\textbf{Game Setting}.
We implement a {\em finite repeated non-cooperative game}, % with perfect information}, (need not to be complete), 
where the strategy parameter $S_i$ of each SC $i$ is the maximum number of VMs shared with other SCs at any given time.
Here, we adapt the concept of \emph{fictitious play} \cite{brown1951iterative}, and assume that each SC does not need to know the utility functions of others.
%The reason of using non-cooperative game is that each SC is selfish, strategic, and not competing for the common resources.
%
SC $i$ determines $S_i$ based on the performance characteristics achieved through sharing with others in the previous round of the game, resulting in a corresponding cost of maintaining the required SLAs.
Algorithm \ref{Algo:noncooperate} describes the details of our non-cooperative repeated game.
In the initial round (without knowledge of other SCs' behavior), each SC makes an initial sharing decision arbitrarily, and begins sharing VMs with other SCs.
Given the solution of the performance model (which takes $\{S_1^{(0)},...,S_K^{(0)}\}$ as input), each SC maximizes its utility, to determine $S_i^{(1)}$, its sharing decision for the next round.
Using its new sharing decision and those from other SCs ($S_j^{(1)},\, \forall j \neq i$) from the previous round, SC $i$ maximizes its utility again, to determine a new sharing decision $S_i^{(2)}$.
This continues until the game converges to an {\em equilibrium point}, as explained next.

\smallskip\noindent\textbf{Analyzing Market Equilibria}. 
A Nash equilibrium point of our proposed repeated game represents the game state in which no SC has any incentive to improve its sharing decision \cite{fudenberg1991game}.
In our work, we are primarily interested in \emph{pure strategy} Nash Equilibria (NE) \cite{fudenberg1991game} as it is more practical to implement and realize for a detailed reasoning. More importantly, we have designed utility functions for the SCs that take as arguments, parameters that are practically relevant to our problem, and are expressions that best reflect SC satisfaction levels. However, in the process, we could not strictly preserve salient mathematical properties related to the utility functions that allow us to derive closed form results about Market Equilibria (ME) from existing seminal works in micro-economic theory, \emph{forcing us to take an experimental stance to characterize equilibria.} Below, we briefly rationalize our stance in the light of the inapplicability of seminal game-theory theorems in characterizing ME in our work. 
\emph{A detailed explanation of our rationale (along with a description of the salient mathematical properties) is in the Appendix \ref{Sec:analyzing_market_detail}}.

\emph{First}, deriving closed form results for our work via the seminal result by \emph{Nash} is not possible due to us (a) dealing with only pure strategy NE, and (b) the utility for an SC might not be \emph{quasi-concave} \cite{boyd2004convex} in general cases. \emph{Second}, deriving closed form results for our work via the seminal result by \emph{Debreu, Fan, and Glicksberg} (derived independently) \cite{debreu1952social,glicksberg1952further,fan1952fixed} in relation to pure strategy NE is not possible due to (a') the quasi-concavity assumption might not always be satisfied (for the peer utility function), which in turn might not guarantee pure strategy NE (violating theorem assumptions), and (b') strategy sets in many applications (including specialized versions of our application setting, i.e., the number of shared VMs is discrete in nature) might not be \emph{continuous} and infinite \cite{fudenberg1991game}, in which case, we would have to go back to using Nash's theorem to guarantee mixed strategy NE (which we do not aim to achieve). Finally, deriving closed form results for our work via the strong seminal result by \emph{Dasgupta and Maskin} \cite{dasgupta1986existence} (that also accounts for discontinuous utility functions) is not possible due to the same reasons in (a) and (b) above.

Despite barriers to closed form analysis, we observe through simulation results (see below) the existence of pure strategy NE for infinite strategy spaces (simulated in a discrete manner, thereby becoming a finite game in simulation), and for \emph{non quasi-concave} SC utility functions. Thus, at least from the experimental results, we observe that for our work, (i) it is not necessary (via the theorem of \emph{Nash}) for quasi concavity to hold for a pure strategy (also discounting the guarantee of only a mixed strategy via Nash's theorem) Nash equilibrium to exist, and (ii) it is not necessary (via the theorem of \emph{Debreu et al.}) for quasi concavity to hold for a pure strategy (also discounting the infinite strategy space assumption via the theorem by \emph{Debreu et al.}, as the simulation is discrete in nature) Nash equilibrium to exist.

\smallskip\noindent\textbf{Reaching Market Equilibria}.
%
%However, in this work, we could not afford a mathematical proof because the quasi-concavity assumption might not always be satisfied (for the SC utility function, which depends on the input from our performance model), which in turn might not guarantee a pure and/or mixed strategy Nash equilibrium \cite{nash1950equilibrium,debreu1952social,glicksberg1952further,fan1952fixed,dasgupta1986existence}.
%
As addressed above, since we could not afford a mathematical proof, in this work, we simulate the game in Algorithm~\ref{Algo:noncooperate} and determine the equilibrium point empirically for a specific price setting ($C_i^P$ and $C_i^G$).
%
%Our proposed repeated game is sensitive to the initialization step (round $0$ in Algorithm~\ref{Algo:noncooperate}), that might lead to multiple Nash equilibrium points due to the linearly separable nature of our SC utility function. 
%
A traditional heuristic to search for one such equilibium point in the game is the numerical \emph{T\^{a}tonnement process} \cite{varian2014intermediate} that is based on the principle of gradient descent. 
%
%However, since in practice the SC strategy elements (e.g., \# of VMs to share) are discrete in nature, we need a \emph{discrete} version of a T\^{a}tonnement process to arrive an equilibrium point. Nonetheless the design and analysis of such a process has been shown to be quite challenging \cite{kaizoji2010multiple}; moreover there is no existing discrete T\^{a}tonnement process to the best of our knowledge. 
%However, due to the discrete nature of  the SC strategy elements (e.g., \# of VMs to share), (i) we need a \emph{discrete} version of a T\^{a}tonnement process to arrive an equilibrium point, (ii) the design and analysis of such a process has been shown to be quite challenging \cite{kaizoji2010multiple}, (iii) there is no existing discrete T\^{a}tonnement process to the best of our knowledge.
In our work, due to the discrete nature of the SC strategy elements (e.g., \# of VMs to share), we need a \emph{discrete} version of a T\^{a}tonnement process to reach an equilibrium point. 
However, the design and analysis of such a process has been shown to be quite challenging \cite{kaizoji2010multiple}; moreover there is no existing discrete T\^{a}tonnement process to the best of our knowledge.
Thus, in our market-based model, we use the non-gradient based \emph{Tabu Search} heuristic \cite{glover1989tabu} to search for an equilibrium value of $S_i^{(r)}$, and reach the global optimum in most cases (by starting at different initial points).

\smallskip\noindent\textbf{Fairness Among SCs}.
A joint social end goal, serving as a benchmark of how well selfish non-cooperative SCs participate in the federation w.r.t. their sharing behavior, is to (a) reach a certain level of fairness (see below for details) among SCs in terms of their utilities, and (b) maximize their individual utilities at ME.
It is important to note here that if we only compare the fairness allocations among SCs, the scenario where all SCs share nothing with others can also be a most fair allocation, but it results in sub-optimal individual utilities (at times an individual utility of zero for the SCs) at ME (see Sect.~\ref{Sec:market_evaluation}).
To achieve our joint social end goal, we need to find a specific price setting (the ratio of $C^G_i$ and $C^P_i$) that enables all SCs to maximize their utilities through sharing VMs while at the same time maintaining an appropriate level of fairness.
In regard to adopting an appropriate fairness measure, we consider in our work the widely popular notion of weighted $\alpha$-fairness \cite{mo2000fair} to combine individual SC  utilities $U_i^{S_i}$ through the function
\begin{equation} \label{Eq:socal_welfare}
W(\alpha, \overrightarrow{S_i}, \overrightarrow{U^{S_i}_i}) = \begin{cases}
\sum_{k=1}^{K} S_i \frac{(U^{S_i}_i)^{1-\alpha}}{1-\alpha} & \alpha \ge 0, \alpha \ne 1;\\
\sum_{k=1}^{K} S_i \log U^{S_i}_i& \alpha = 1.
\end{cases}
\end{equation}
Here, $S_i$, the maximum number of shared VMs, is the weight used to combine the $\alpha$-fairness metric of each SC~$i$, while the parameter $\alpha$ controls the fairness of utility allocations among SCs.
In this work, we evaluate three popular $\alpha$-fairness utility functions, achieving different trade-offs between fairness and economic efficiency: (i) $\alpha = 0$, which gives the utilitarian function \cite{mas1995microeconomic} (denoting minimum fairness), (ii) $\alpha = \infty$, which results in max-min fairness, and (iii) $\alpha = 1$, which gives proportional fairness.
For each fairness function defined by $\alpha$, our goal is to find the best price setting that motivates SCs, based on their system loads, to participate in the federation and share more of their VMs, i.e., thereby achieving higher values of $\alpha$-fair functions.
We assume that SCs always report the true decisions and utility without releasing detailed information. (The design of an economic mechanism to enforce truthful communication between SCs is beyond the scope here.)

%
%For our cooperative setting, some relevant social welfare (SW) functions include (i) sum of the utilities of different SCs, i.e., the utilitarian function, (ii) \emph{max-min} fair utility allocations among SCs, and (iii) a \emph{proportional fair} utility allocation among SCs. 
%These functions are special cases of the weighted $\alpha$-fair welfare functions \cite{mo2000fair} given by: 
%Here, we use the notion of the weighted $\alpha$-fair welfare functions  to define our social welfare function:

%
%Smaller values of $\alpha$ mean the social welfare focuses on the summation, and larger values of $\alpha$ mean the social welfare favors the fairness. 
%

%
%Moreover, we use the number of maximum shared VMs, $S_i$, as weights to force the higher social welfare happening at the case, where an SC that is willing to contribute more VMs has higher utility.
\section{Evaluation and Validation} \label{Sec:eval}

We first validate the accuracy of our performance model, the results of which are needed as input parameters to the market-based model.
To this end, we compute the solution of our approximate model (in Sect.~\ref{Sec:performance_model}) numerically, and compare it to the solution of the exact model (computed through a C++-based simulator).
%$\mathcal{M}$, in steady state.
%
We then use our market-based model to investigate \emph{how the price of using shared VMs from other SCs affects achieving higher summation of weighted utilities}.

%\emph{investigate the increase/decrease of social welfare with the increase in price of using shared VMs}. 

%To evaluate the market-based model, we implement the game described in
%Algorithm \ref{Algo:noncooperate}, and analyze the incentive for SCs to participate in the cooperative when SCs in the cooperative have different combinations of utilizations.
%
%The implementation converges to an equilibrium point within (no more than) $10$ rounds.
%
%%%\leana{What about how the market-based model is computed?  A different simulation?}
%
%\emph{Our goal is to investigate the increase/decrease of cooperative efficiency with the increase in price of using shared VMs.}

\subsection{Performance model validation}

%Since the market-based model use performance parameters computed from our performance model, we first validate the accuracy of our performance model.

\begin{figure}[tp]
	\captionsetup{farskip=0pt}
	\centering
	\subfloat[][$10$ VMs]{\includegraphics[width=0.24\textwidth]{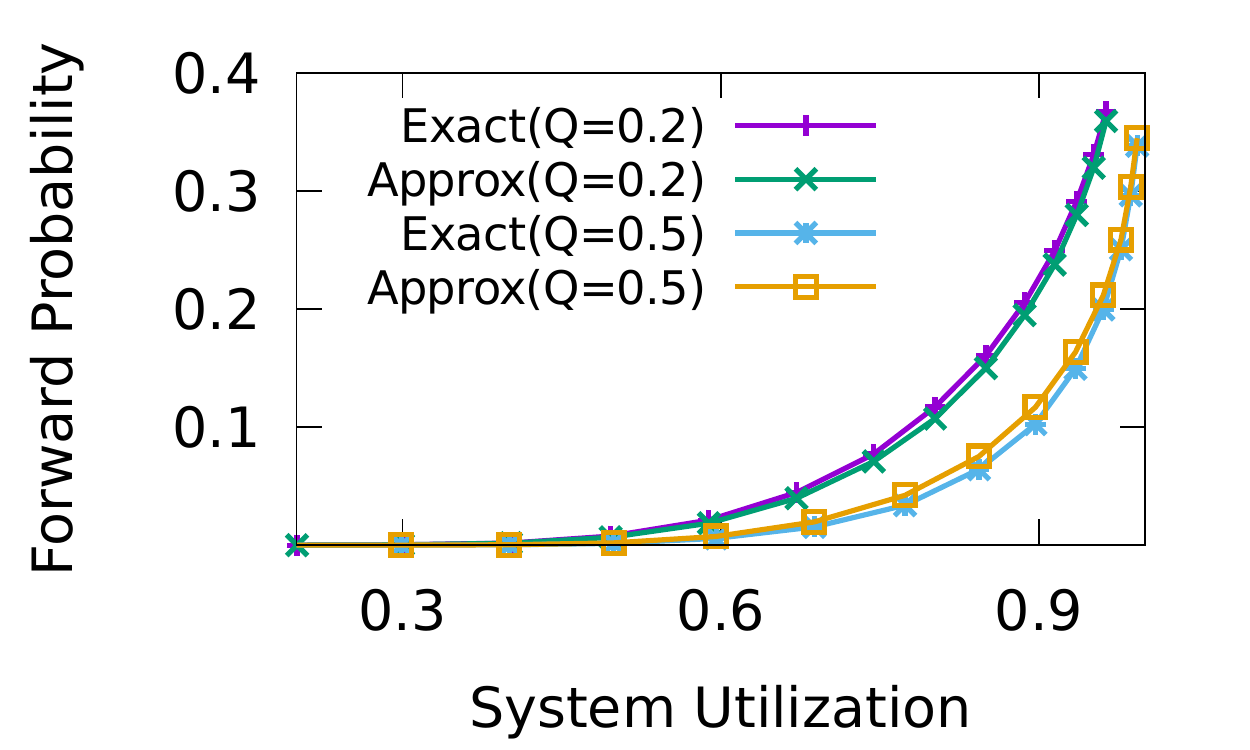}\label{Fig:forward_prob_10}}
	\subfloat[][$100$ VMs]{\includegraphics[width=0.24\textwidth]{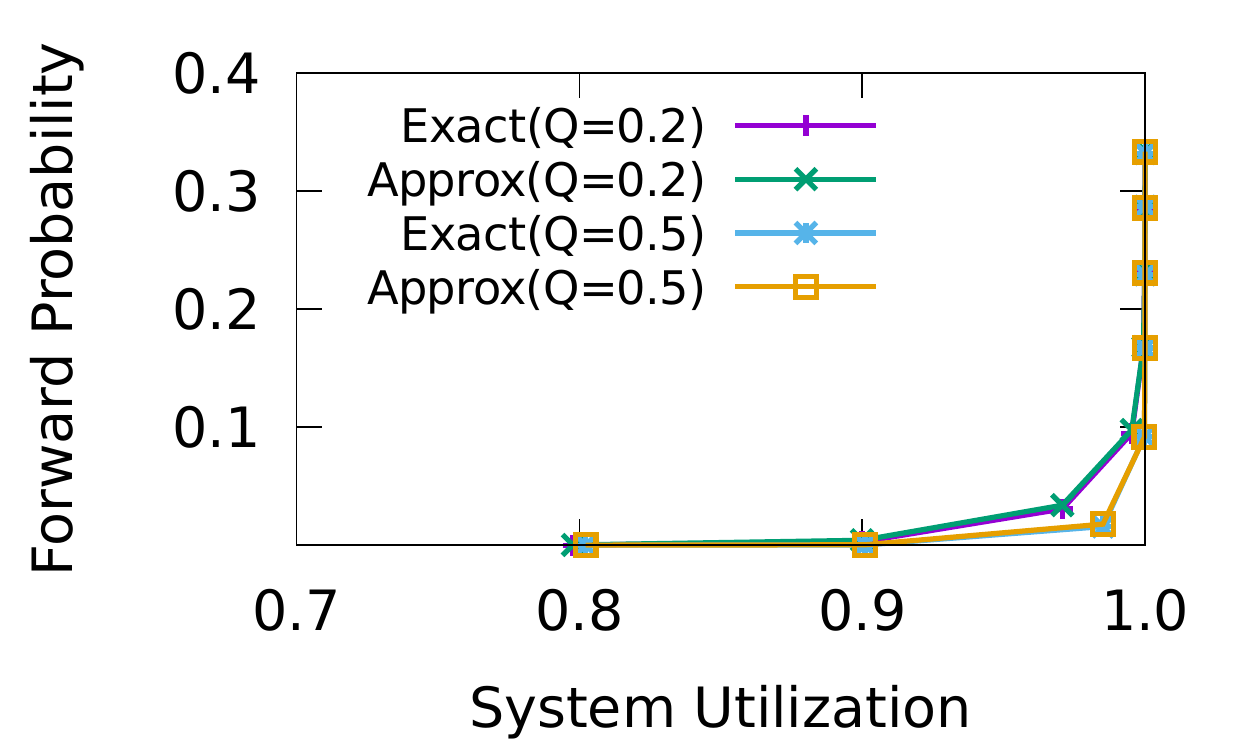}\label{Fig:forward_prob_100}}
	\caption{Comparing the result of forwarding estimation in $10$ and $100$ VMs with QoS = $0.2$ and $0.5$.}
	\vspace{-1em}
\label{Fig:forward_result_10}
\end{figure}

\smallskip\noindent\textbf{SC without Sharing Resources}.
Here, we start with the accuracy evaluation of our forward probability estimation in Sect.~\ref{Sec:forward_model}, since this is a measure used by all other models.
Moreover, to demonstrate that SCs have more incentives to participate in the federation, we compare the results of two clouds, which have $10$ and $100$ VMs respectively, with the SLAs of $Q_i=0.2$ and $Q_i=0.5$ under various Poisson arrival rates; each request has an exponential service time with rate $1$.
In order to correctly compare the results among two SCs, in Fig. \ref{Fig:forward_result_10}, we show the estimated forward probability under different system utilizations (by increasing the arrival rate).
As shown in the figure, for both clouds, the probability of forwarding is higher for smaller QoS values, and our estimation properly predicts the forward probability under different settings. 
It is easy to see that the cloud with fewer VMs has higher forwarding probability under the same system utilization.
Thus, if an SC does not want to increase its investments in infrastructure, it needs some mechanism to decrease its forwarding probability to reduce the cost of satisfying SLAs.
In the following experiments, each SC in the federation has $10$ VMs by default with exponential service time with rate $\mu=1$ and QoS $Q_i=0.2$.

%\begin{figure}[tp]
%	\centering
%	\begin{subfigure}[b]{0.48\linewidth}
%		\includegraphics[width=1\textwidth]{figure/forward_public_10_prob.pdf}
%		\caption{$10$ VMs}
%		\label{Fig:forward_prob_10}
%	\end{subfigure}
%	~ %add desired spacing between images, e. g. ~, \quad, \qquad, \hfill etc. 
%	\begin{subfigure}[b]{0.48\linewidth}
%		\includegraphics[width=1\textwidth]{figure/forward_public_100_prob.pdf}
%		\caption{$100$ VMs}
%		\label{Fig:forward_prob_100}
%	\end{subfigure}
%	\caption{Comparing the result of forwarding estimation in $10$ and $100$ VMs with QoS = $0.2$ and $0.5$.}
%	\label{Fig:forward_result_10}
%\end{figure}

\medskip\noindent\textbf{Approximate Model}.
%
%\sunghan{I still work on improving the model since I completely change the model in the weekend to reduce computation complexity.}
%
In this section, we performed extensive experiments to validate the accuracy of the approximate model presented in Sect.~\ref{Sec:approximate_model}.
Here, we want to investigate how well our approximate model performs as a function of the different number of shared VMs and system utilizations.

\begin{figure}[tp]
	\captionsetup{farskip=0pt}
	\centering
	\subfloat[][$\overline{I_i^{S_i}}$, $2$ SCs, $10$ VMs]{\includegraphics[width=0.24\textwidth]{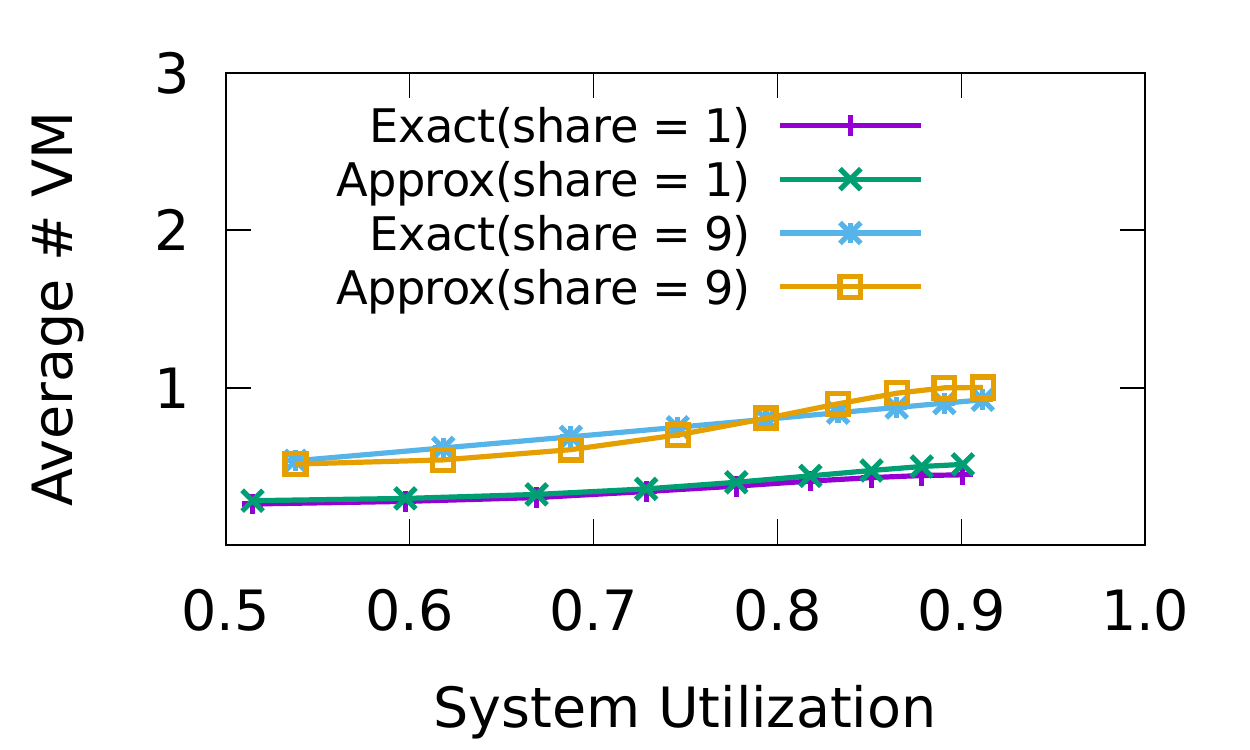}\label{Fig:model_share_in_2}}
	\subfloat[][$\overline{O_i^{S_i}}$, $2$ SCs, $10$ VMs]{\includegraphics[width=0.24\textwidth]{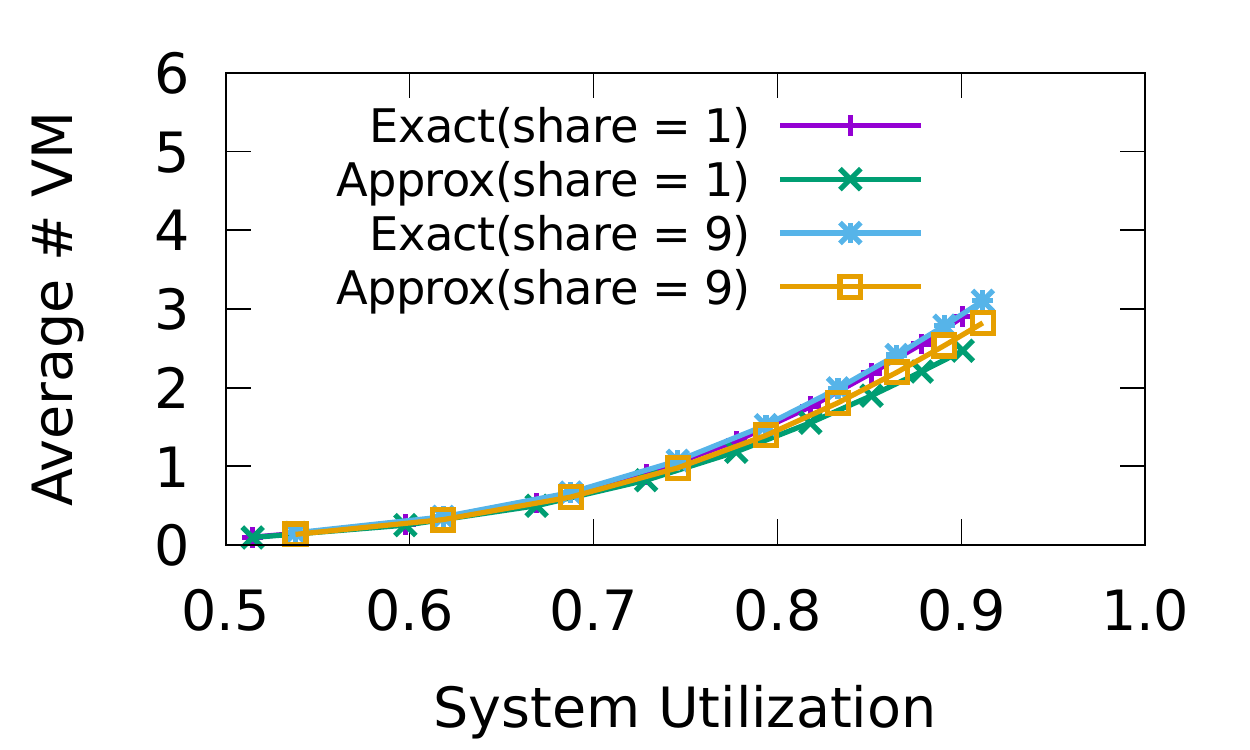}\label{Fig:model_share_out_2}}
	\qquad
	\subfloat[][$\overline{I_i^{S_i}}$, $10$ SCs, $10$ VMs]{\includegraphics[width=0.24\textwidth]{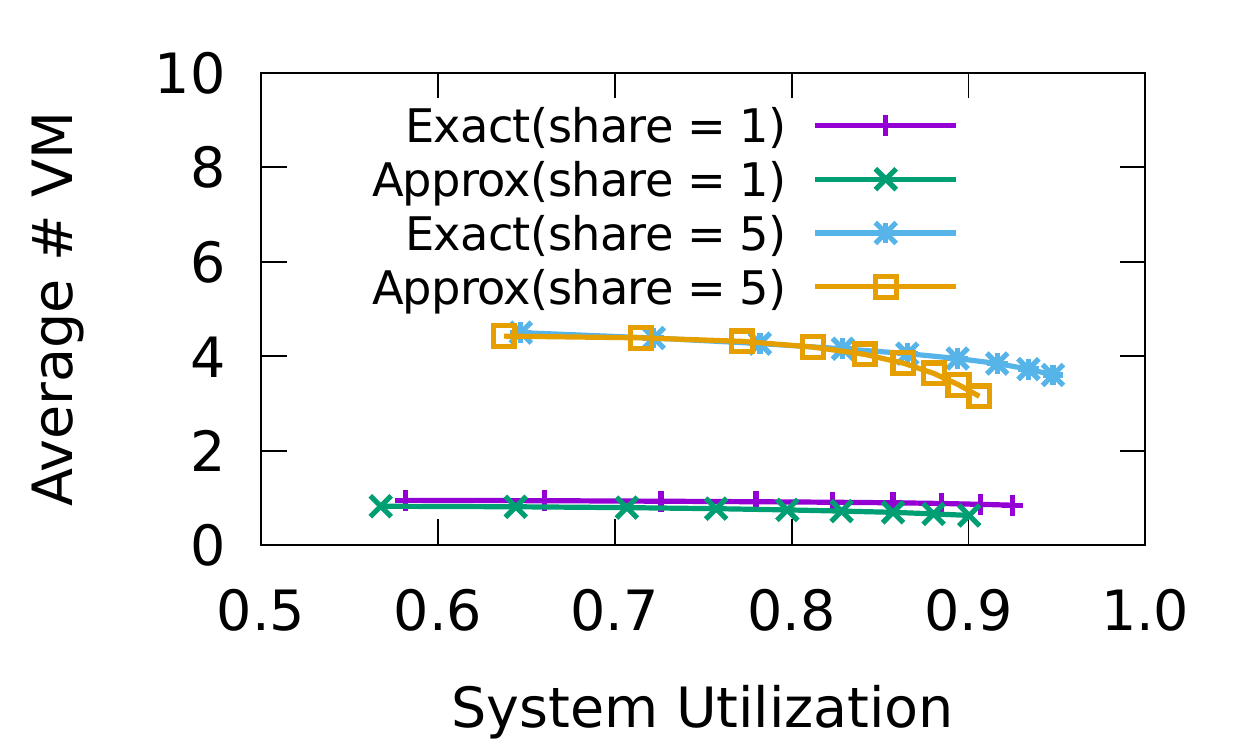}\label{Fig:model_share_in_10}}
	\subfloat[][$\overline{O_i^{S_i}}$, $10$ SCs, $10$ VMs]{\includegraphics[width=0.24\textwidth]{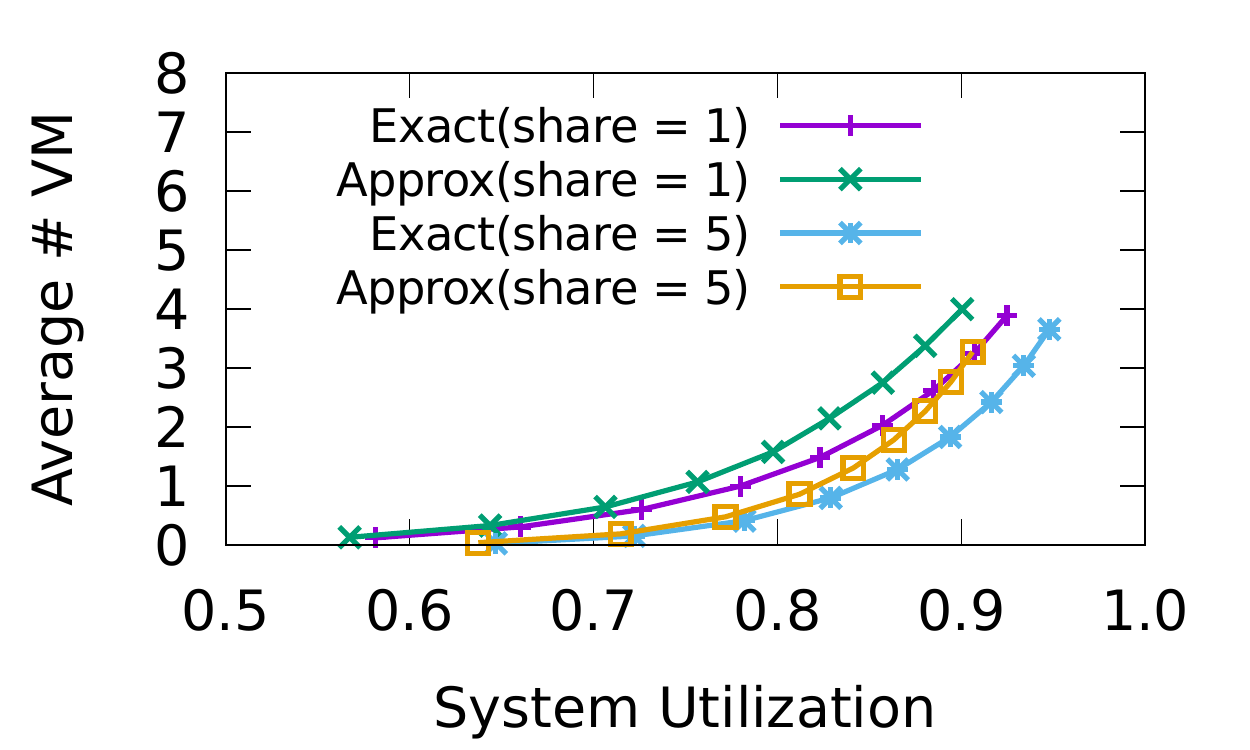}\label{Fig:model_share_out_10}}
	\qquad
	\subfloat[][$\overline{I_i^{S_i}}$, $2$ SCs, $100$ VMs]{\includegraphics[width=0.24\textwidth]{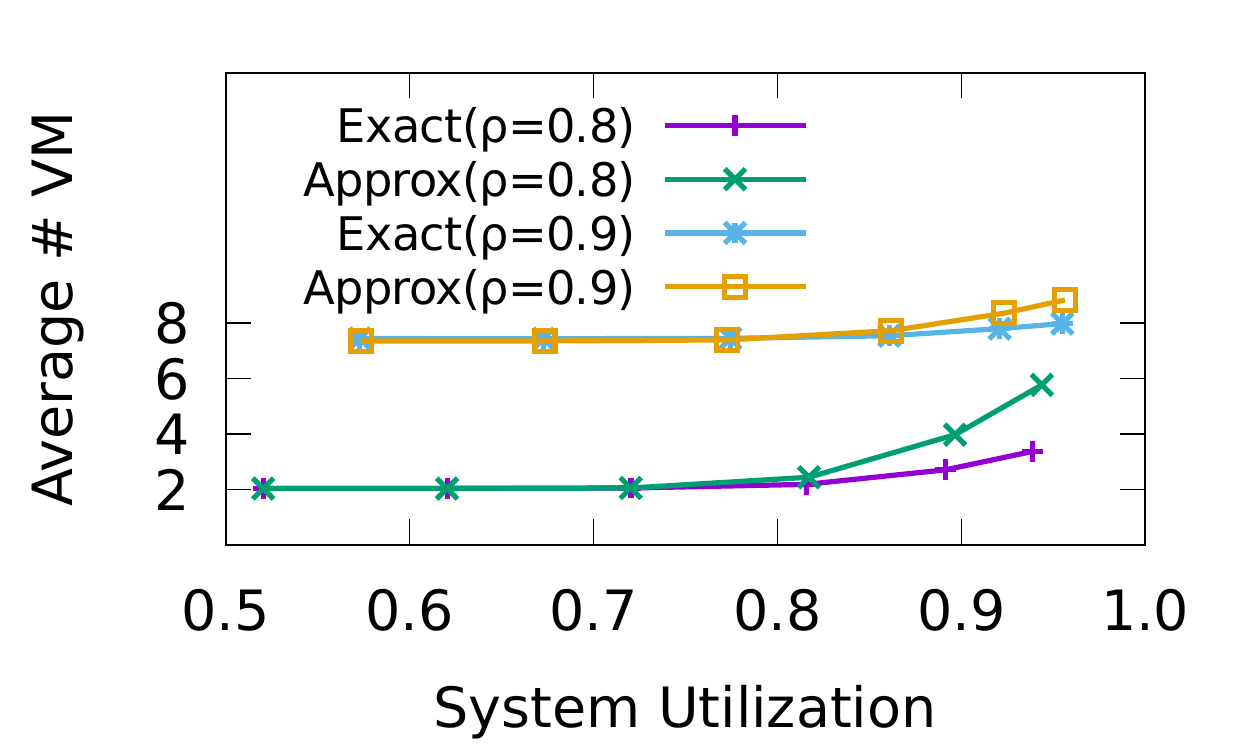}\label{Fig:model_share_in_2_100}}
	\subfloat[][$\overline{O_i^{S_i}}$, $2$ SCs, $100$ VMs]{\includegraphics[width=0.24\textwidth]{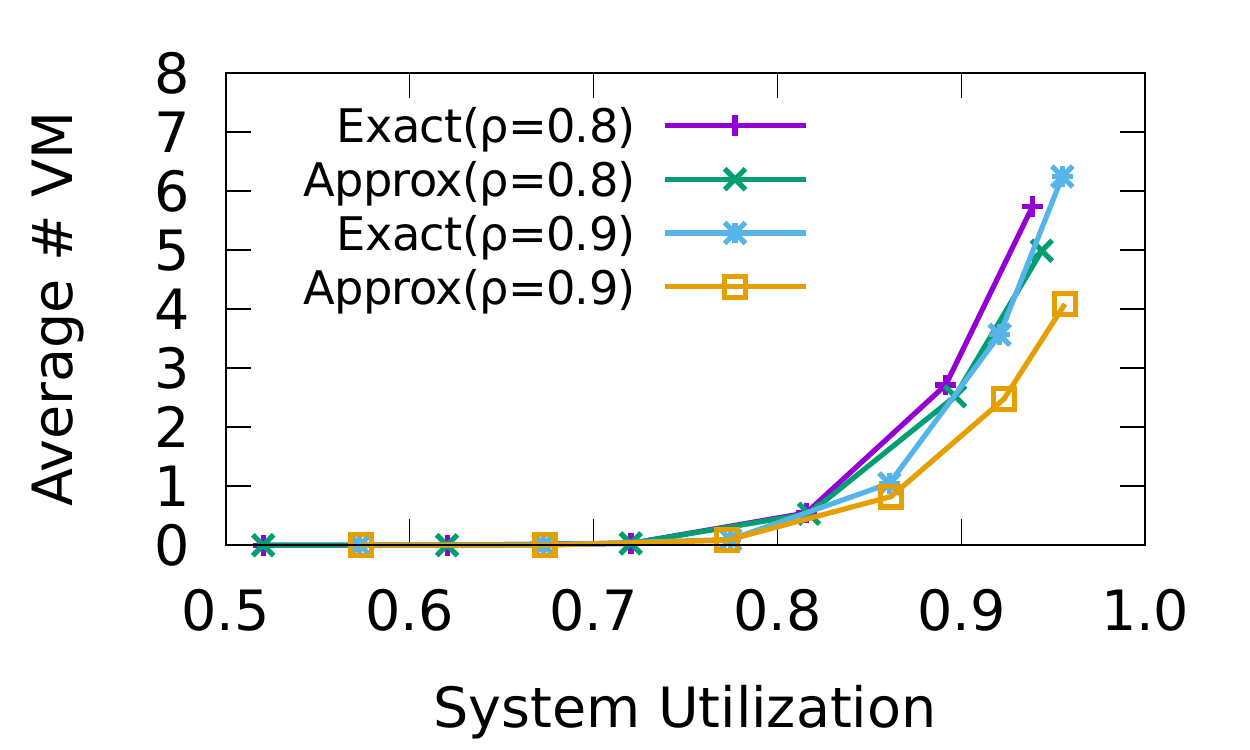}\label{Fig:model_share_out_2_100}}
	\caption{Validating approx.\ perf.\ model (2 SCs and 10 SCs)}
	\vspace{-1em}
	\label{Fig:model_share}
\end{figure}

We begin with a 2-SC federation scenario.
We fix the arrival rate of one SC to $7$ and the number of shared VMs to $5$ (total $10$ VMs), and vary the number of shared VMs and system load (by changing the arrival rate) of another SC, referred to as target SC.
Figures \ref{Fig:model_share_in_2} and \ref{Fig:model_share_out_2} illustrate the performance metrics of interest when the target SC shares $1$ and $9$ VM(s) under different system loads.
(Due to lack of space, we omit $\overline{P_i^{S_i}}$ as its estimation remains accurate.)
As shown in the figures, the exact and approximate $\overline{I_i^{S_i}}$ and $\overline{O_i^{S_i}}$ are nearly the same when the target SC shares very few VMs.
The inaccuracy of our approximate model grows when the target SC shares more VMs (as compared to a scenario with $1$ shared VM), but is still within $10\%$. 
Thus, the difference between $\overline{I_i^{S_i}}$ and $\overline{O_i^{S_i}}$ (see Eq. (\ref{Eq:cost})) remains accurate (within $10\%$ of the exact solution). 
%
%Since our market-based model only requires (as input) the difference between these metrics (see Eq. \ref{Eq:cost}), this is still a sufficiently accurate approximation.

%
%Moreover, Figure \ref{Fig:model_share_p_2} shows that our approximate model properly estimates $\overline{S_i^{P}}$ no matter how many VMs shared by SC $1$.
We now illustrate how the approximation error grows in larger systems.
Firstly, we consider a 10-SC federation scenario \sunghan{(each with a total of $10$ VMs)}, and fix $9$ SCs' settings, which have the following number of shared VMs $(3,3,3,2,2,2,1,1,1)$, and corresponding arrival rate  $(7,7,7,8,8,8,9,9,9)$. 
Figures \ref{Fig:model_share_in_10} and \ref{Fig:model_share_out_10} illustrate the performance metrics of interest when the target SC shares $1$ and $5$ VM(s) under different system loads.
We still observe that the difference between the exact and approximate $\overline{I_i^{S_i}}$ and $\overline{O_i^{S_i}}$ remains small (within $10\%$ of the exact solution) when the system utilization is lower than $0.8$ (within $20\%$ when the system utilization is lower than $0.9$). 
%
%When the number of shared VMs increases, the inaccuracy of our approximation grows but not significantly (especially when the system utilization is lower than $0.8$).
%
%Unlike the results in the 2-SC scenario, where the results of the approximate model are under-estimated, the results of the approximate model here are over-estimated. 
%
Generally, we can observe that the results of approximated $\overline{I_i^{S_i}}$ are under-estimated when the system has very high utilization because our approximate model breaks the direct relationship between the target SC and all other SCs (we only consider the connection between SC $i$ and SC $i-1$); thus, the target SC might under-estimate the number of queued requests at all other SCs.
For the same reason, the results of approximated $\overline{O_i^{S_i}}$ are over-estimated.
%, and we aggressively predict the value of $\overline{I_i^{S_i}}$, resulting in over-estimate of metrics.
%
However, the difference between $\overline{I_i^{S_i}}$ and $\overline{O_i^{S_i}}$ remains accurate (within $20\%$ of the exact solution) when the system utilization is lower than $0.9$.
Second, we consider again a 2-SC federation scenario, with $100$ VMs per SC.
We fix the the number of shared VMs at $10$ for both SCs, and vary system load for both of them.
Figures \ref{Fig:model_share_in_2_100} and \ref{Fig:model_share_out_2_100} illustrate the performance metrics of interest when one SC has system utilization of $0.8$ and $0.9$ under different system loads of the target SC.
We still observe that the difference between $\overline{I_i^{S_i}}$ and $\overline{O_i^{S_i}}$ remains accurate (within $20\%$ of the exact solution) when the system utilization of the target SC is lower than $0.9$.

%our model is designed to process high system utilization.

\medskip\noindent\textbf{Summary}.
Our extensive experiments indicate that our approximate model estimates $\overline{I_i^{S_i}}$ and $\overline{O_i^{S_i}}$ within $20\%$ of the exact solution, under a variety of scenarios, while saving significant computation time.
More importantly, the accuracy of the {\em difference\/} between $\overline{I_i^{S_i}}$ and $\overline{O_i^{S_i}}$, and $\overline{P_i^{S_i}}$, which are the parameters needed by the market-based model, are within $10\%$ of the exact solution when the system utilization is reasonable.
Overall, we believe that our approximate model is useful in estimating performance characteristics of the federation, as needed in the market-based model.

\subsection{Market-based Model Evaluation} \label{Sec:market_evaluation}

\begin{figure}[tp]
	\captionsetup{farskip=0pt}
	\centering
	\subfloat[][All SCs with ${UF}_0$]{\includegraphics[width=0.24\textwidth]{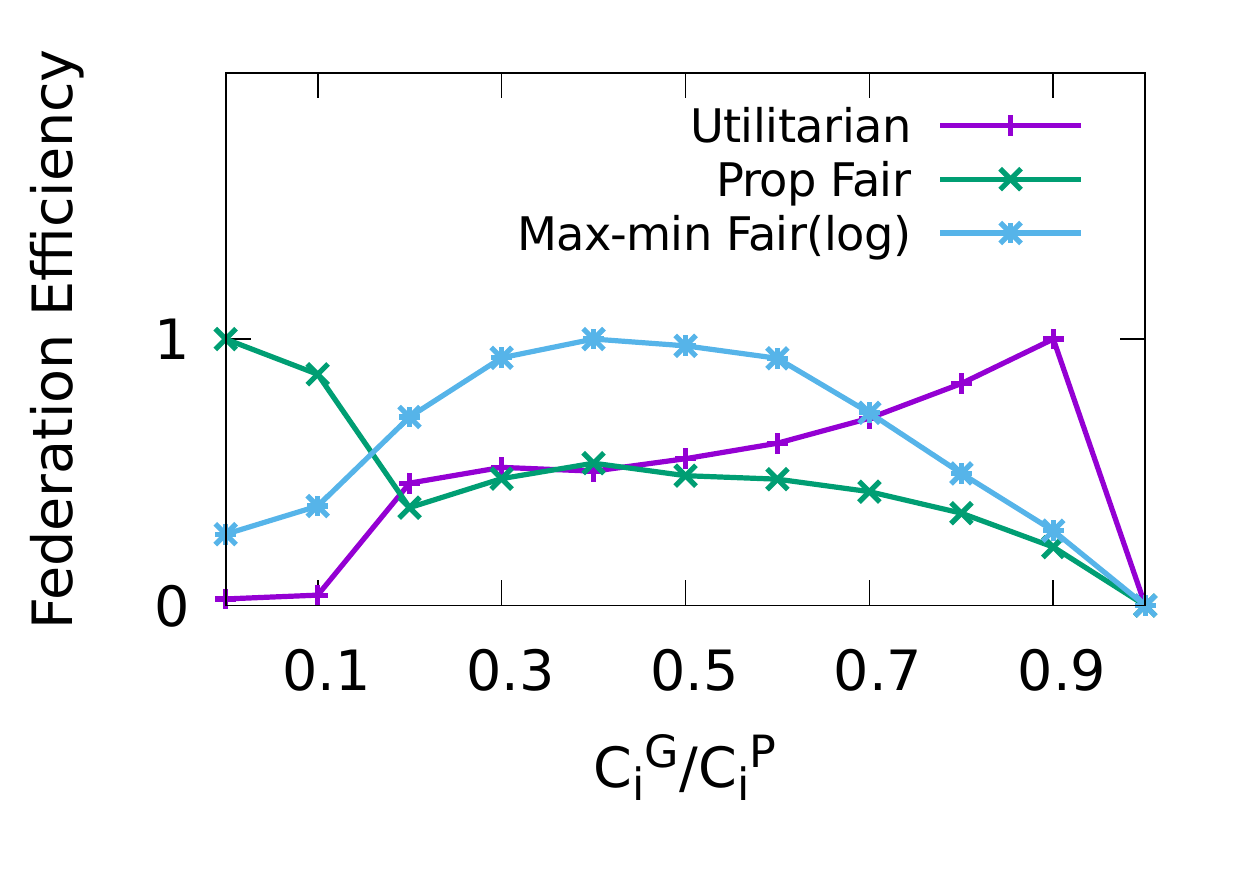}\label{Fig:market_3_6810}}
	\subfloat[][All SCs with ${UF}_1$]{\includegraphics[width=0.24\textwidth]{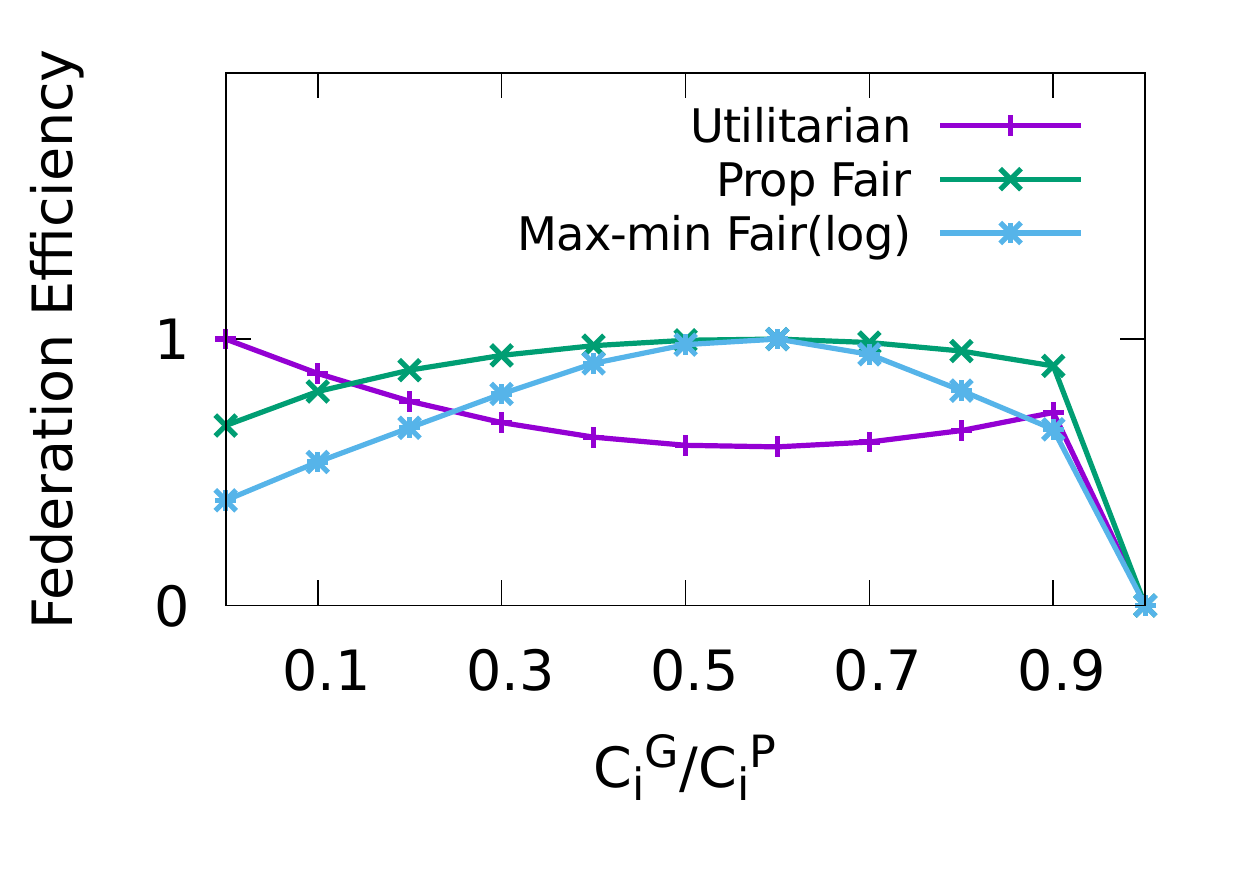}\label{Fig:market_3_6810_all_util}}
	\qquad
	\subfloat[][All SCs with ${UF}_0$]{\includegraphics[width=0.24\textwidth]{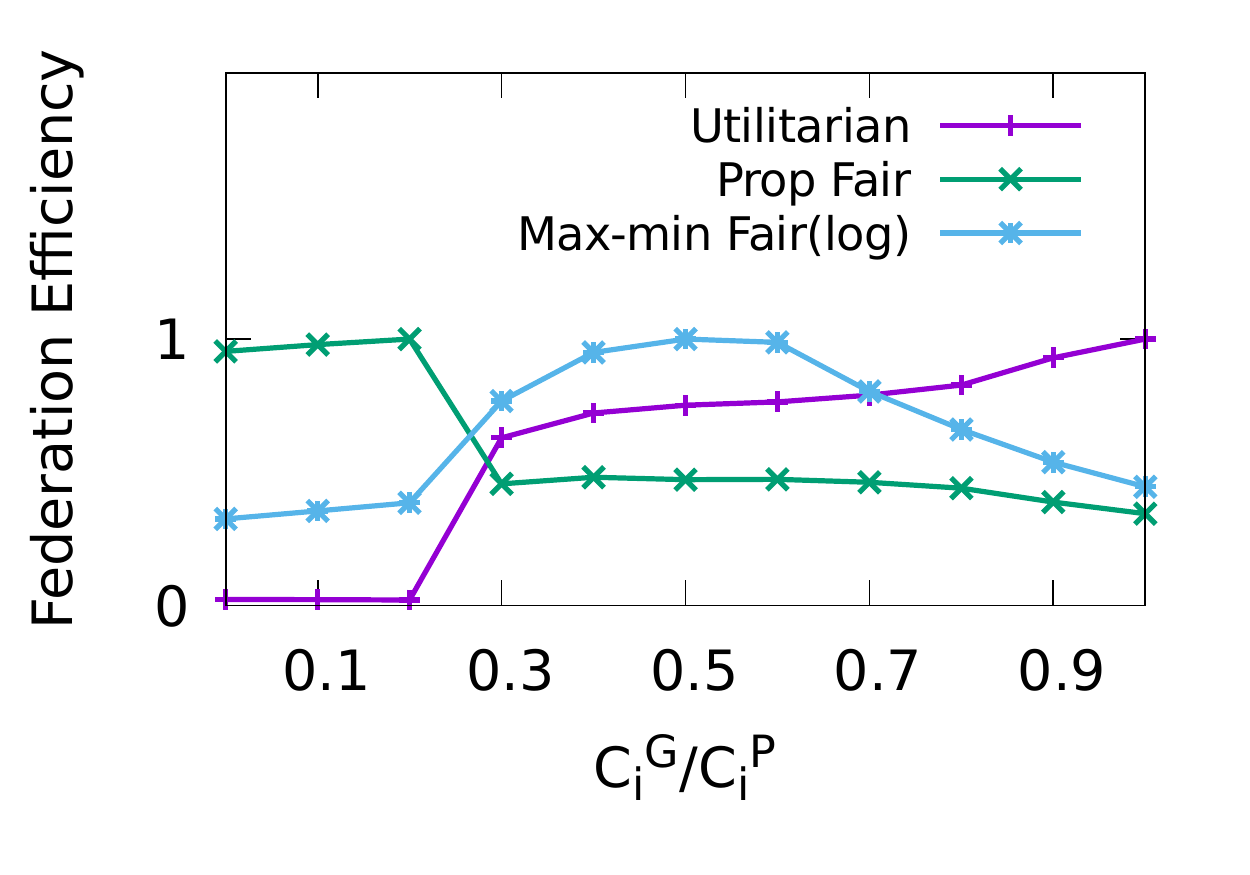}\label{Fig:market_3_8910}}
	\subfloat[][All SCs with ${UF}_1$]{\includegraphics[width=0.24\textwidth]{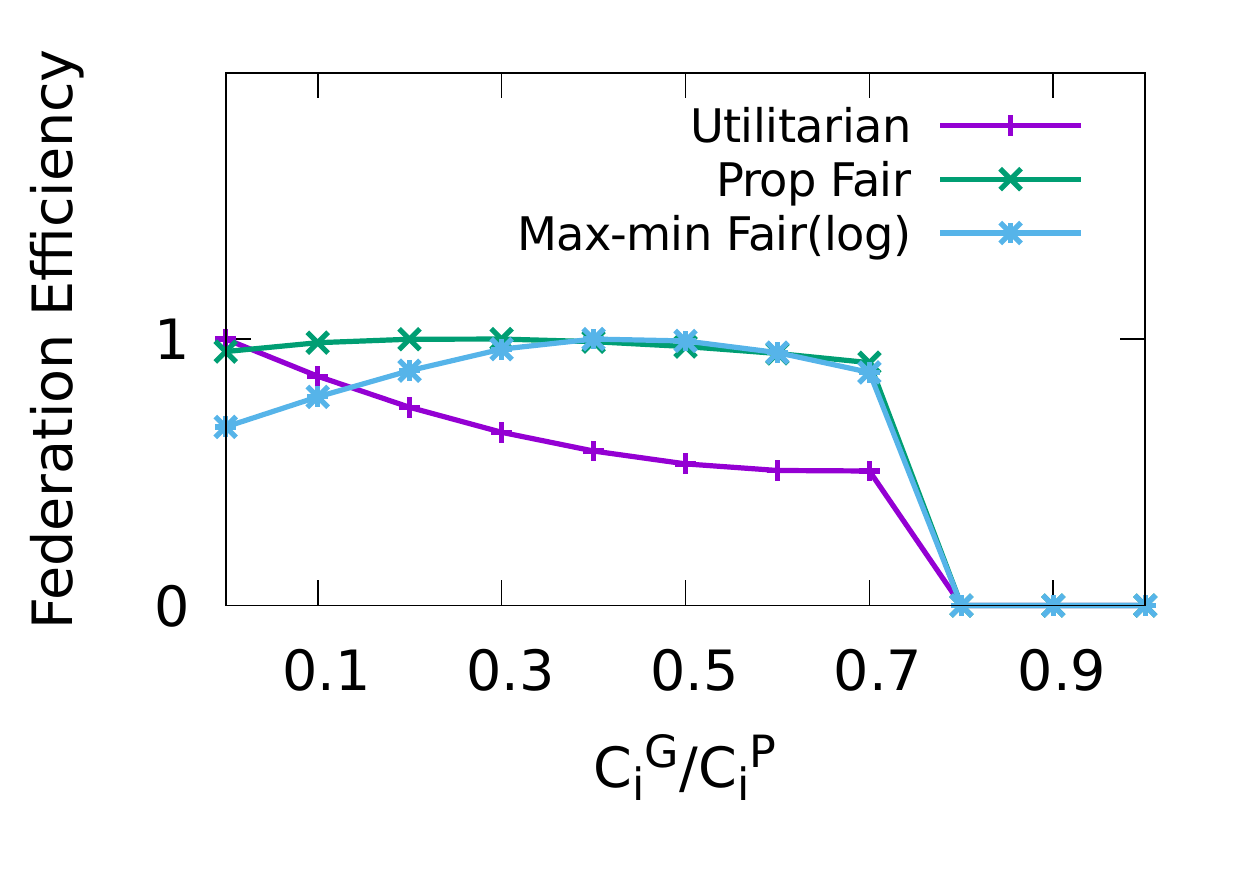}\label{Fig:market_3_567_all_util}}
%	\qquad
%	\subfloat[][only SC $\rho=0.55$ with ${UF}_1$]{\includegraphics[width=0.24\textwidth]{figure/market_3_6810_util.pdf}\label{Fig:market_3_6810_util}}
%	\subfloat[][All SCs with ${UF}_0$]{\includegraphics[width=0.24\textwidth]{figure/market_3_567.pdf}\label{Fig:market_3_567}}
	\caption{Market results in 3-SC scenarios: (a)(b) are results where $3$ SCs have $\rho_i=0.58, 0.73, 0.84$, (c) is the result where $3$ SCs have $\rho_i=0.73, 0.79, 0.84$, (d) are the results where $3$ SCs have $\rho_i=0.49, 0.58, 0.66$}
	\vspace{-1em}
	\label{Fig:market_result_3}
\end{figure}

Here, we perform experiments to investigate how $\frac{C_{i}^{G}}{C_{i}^{P}}$, $U^{S_i}_i$, and $W(\alpha, \overrightarrow{S_i}, \overrightarrow{U^{S_i}_i})$, affect the criteria for SCs to participate in the federation.
\sunghan{Due to lack of space, we focus on evaluating 3-SC scenarios(in Fig.~\ref{Fig:market_result_3}), where each SC has $10$ VMs (as a representative example) in the evaluation, to better explain the effects of system utilizations on the game model; results for other SC-scenarios are qualitatively similar.}
Here, we display the ratio of the achieved value of the W metric (see Sect~\ref{Sec:non_cooperative}) to the (empirical) market efficient value of the W metric, as a measure of \emph{federation efficiency}, for a given mixture of SC utility functions. %For clarity of presentation, we depict the logarithmic values of the max-min W metric.
If no SCs are willing to participate in the federation, we depict it as zero federation efficiency (since the value of the W metric is always greater than zero).
%
%Moreover, in figures, we plot the social welfare function with $\gamma=0$ even when SCs use marginal cost reduction as its utility function; thus, enable us to compare the results with other figures.
%

%In the first set of experiments, 
We first consider scenarios where the $3$ SCs have significantly \emph{different system loads} ($\rho_i=0.58, 0.73, 0.84$).
Fig. \ref{Fig:market_3_6810} illustrates the case where all SCs choose $UF_{0}$ ($\gamma=0$) as their utility function; Fig. \ref{Fig:market_3_6810_all_util} illustrates the case where all SCs choose $UF_{1}$ ($\gamma=1$) as their utility function.
%.
As shown in the figures, if all SCs chooses ${UF}_0$, the utilitarian W metric increases with increase in $C_{i}^{G}/C_{i}^{P}$ (except when $C_{i}^{G}/C_{i}^{P}$ is nearing $1$), since the SCs choosing $UF_{0}$ as their utility are incentivized to share more VMs to reduce their net cost. 
When $C_{i}^{G}/C_{i}^{P}$ is nearing $1$, the federation cannot be formed because SCs with high utilizations do not reduce cost through using shared VMs, compared to that when resorting to a public cloud, and low utilization SCs do not generate enough demand to make high utilization SCs remain profitable. 
If all SCs use ${UF}_1$, they would only share $1$ VM with others even when $C_{i}^{G}/C_{i}^{P}$ increases because, in our setting, the increase in marginal cost reduction with increase in number of shared VMs is not sufficient to encourage SCs to contribute more VMs. 
Moreover, since all SCs only shared $1$ VM when they use ${UF}_1$, both proportional W metric and max-min W metric achieve the same maximum state (due to the same weight for all SCs in Eq. (\ref{Eq:socal_welfare})), as shown in Fig. \ref{Fig:market_3_6810_all_util}.
In other cases, the results of the proportional W metric depend on the behavior of the lower utilization SCs.
If these SCs choose ${UF}_0$, their cost reductions with increase in number of shared VMs are greater than high utilization SCs; thus, the maximum proportional W metric can only happen when all SCs share few VMs. %, i.e., $\frac{C_{i}^{G}}{C_{i}^{P}} = 0$.
%
%However, if lower utilization SCs choose ${UF}_1$, which limits their number of shared VMs, high utilization SCs can increase their shared VMs to achieve higher cost reduction, resulting in better proportional SW, as shown in Fig. \ref{Fig:market_3_6810_util}.

In Fig. \ref{Fig:market_3_8910}, we consider scenarios where $3$ SCs have similarly \emph{high system loads} ($\rho_i=0.73, 0.79, 0.84$), where all of them consider ${UF}_0$.
In this scenario, the results are similar to the cases in Fig. \ref{Fig:market_3_6810}; however, unlike the scenario where SCs having significant different utilizations \emph{are not incentivized} to join the federation when $C_{i}^{G}/C_{i}^{P} = 1$, SCs in a scenario when they have similar high utilizations, are \emph{incentivized} to cooperate when $C_{i}^{G}/C_{i}^{P} = 1$.  
This is because high utilization SCs share similar number of VMs with each other, resulting in canceling out the cost of using shared VMs. 
In Fig. \ref{Fig:market_3_567_all_util}, we consider scenarios where $3$ SCs have similarly \emph{median system loads} ($\rho_i=0.49, 0.58, 0.66$), where all of them consider ${UF}_1$.
The results in these scenarios are similar to what we have discussed above, however, we observe the federation cannot be formed when $C_{i}^{G}/C_{i}^{P}$ is beyond $\approx 0.8$.
This is because all low utilization SCs do not generate enough revenue from their incoming VM demand from other SCs to offset their costs of using shared VMs from other SCs. 
%would rather queue requests and forward the requests to a public cloud occasionally than use shared VMs immediately when $\frac{C_{i}^{G}}{C_{i}^{P}}$ is more than a threshold.

\medskip\noindent\textbf{Summary}.
Our extensive experimental evaluation indicates \emph{three} $C_{i}^{G}/C_{i}^{P}$ regions of operation to maximize various W metrics. 
When \emph{maximizing proportional fairness based W metric} is the goal of the federation, the value of $C_{i}^{G}/C_{i}^{P}$ should be set in the \emph{lower range} of $C_{i}^{G}/C_{i}^{P}$ (between $0$ and $0.3$ in our example setting). 
When \emph{maximizing max-min fairness based W metric} is the goal, the value of $C_{i}^{G}/C_{i}^{P}$ should be set in the \emph{middle range} of $C_{i}^{G}/C_{i}^{P}$ (between $0.3$ and $0.7$ in our example setting). 
Finally, when \emph{maximizing utilitarian W metric} is the goal, the value of $C_{i}^{G}/C_{i}^{P}$ should be set in the \emph{high range} of $C_{i}^{G}/C_{i}^{P}$ (between $0.7$ and $1$ in our example setting). 
However, the utilitarian setting also runs the risk of breaking the federation at a certain high value of $C_{i}^{G}/C_{i}^{P}$ at which no SC would be willing to cooperate.

\subsection{Computational Overhead}

\sunghan{Here, we discuss the cost of computing our performance model and the market-based model.}

\begin{figure}[tp]
	\captionsetup{farskip=0pt}
	\centering
	\subfloat[][Approximate performance model]{\includegraphics[width=0.24\textwidth]{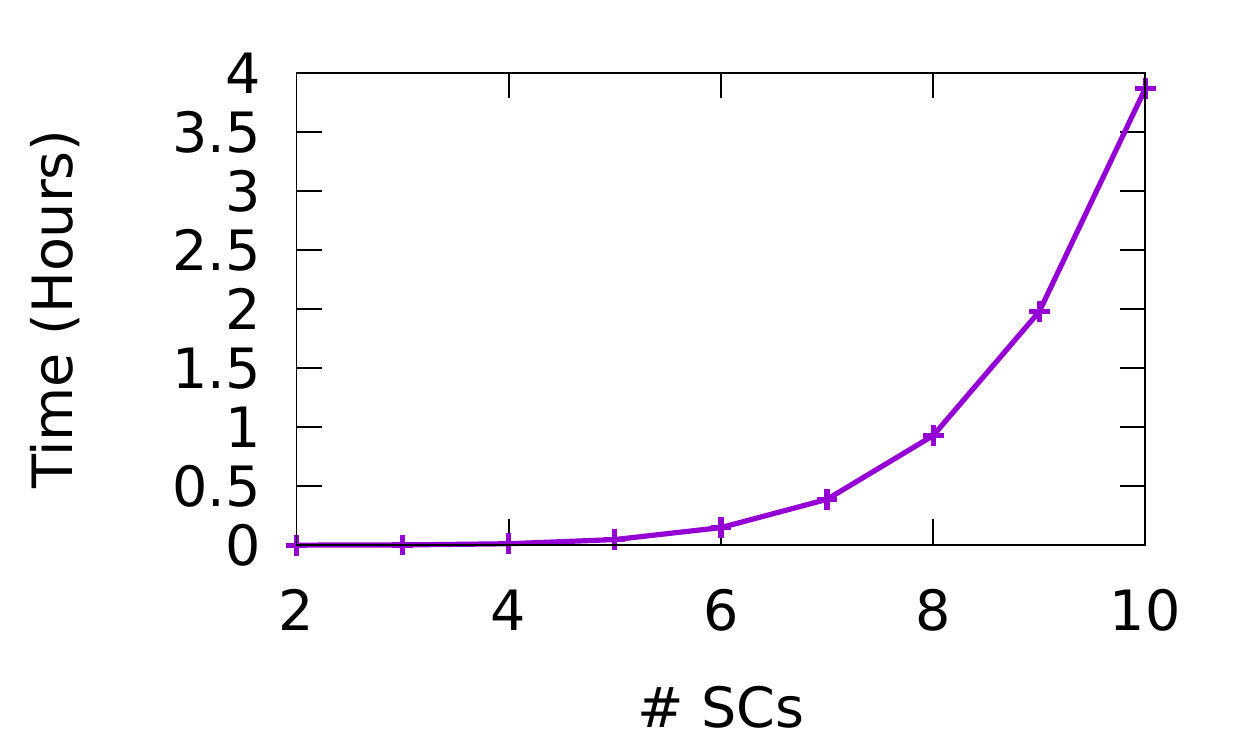}\label{Fig:time_complex_performance}}
	\subfloat[][Game model]{\includegraphics[width=0.24\textwidth]{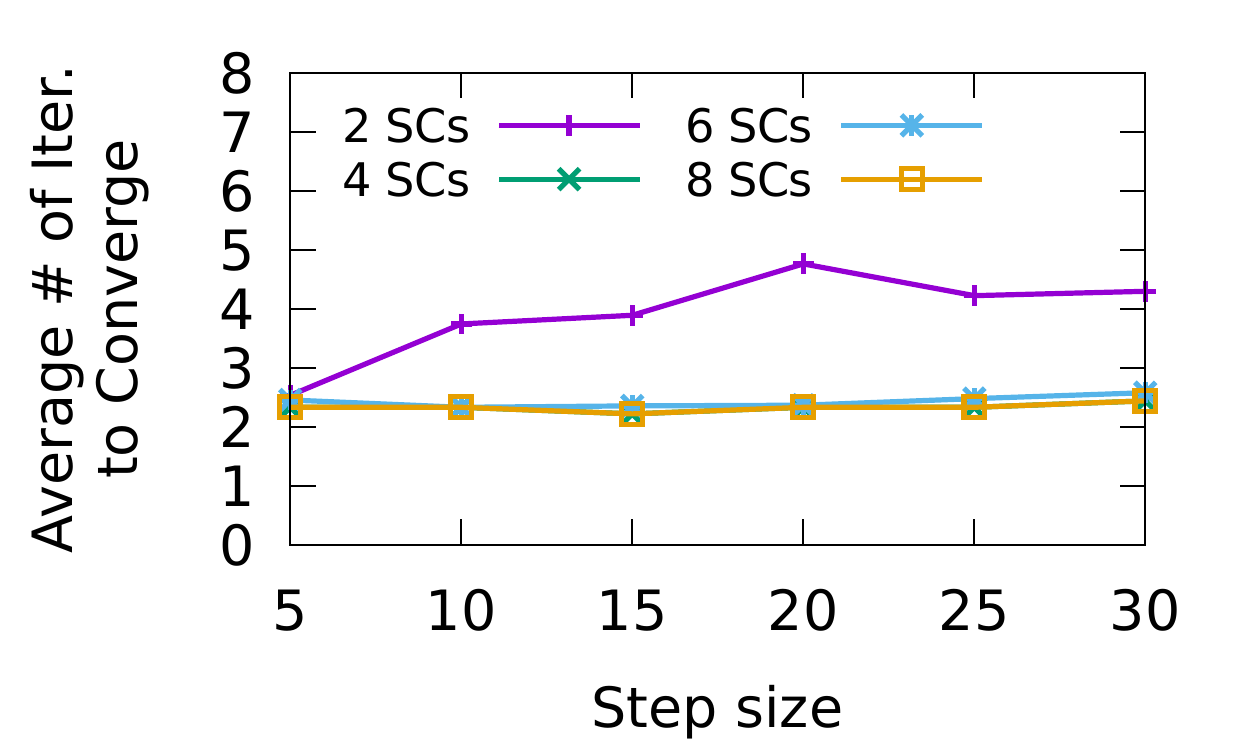}\label{Fig:time_complex_game}}
	\caption{Time complexity of the performance model and the game model}
	\vspace{-1em}
	\label{Fig:time_complex}
\end{figure}

\sunghan{\noindent\textbf{Performance Model: }Our approximate model can significantly reduce the state space of the Markov model (see Sect.~\ref{Sec:approximate_model}). For instance, in a 10-SC scenario with each SC sharing $5$ VMs, the detailed model has \emph{$\approx 9$-billion} states, whose generation and solution requires a substantial amount of space and computation time.
However, our approximate model only needs to build ten Markov models with \emph{$\approx 1$-million} states each, and solve the corresponding matrices.
Fig. \ref{Fig:time_complex_performance} illustrates the computation time of the approximate model with $2-10$ SCs, each with $10$ VMs and sharing $2$ VMs. We observe that the computation time increases with the number of SCs due to generating and solving larger matrices. Since our approximate model significantly reduces the state space, it can estimate the results faster and with less memory.}

\sunghan{\noindent\textbf{Market-based Model: }SCs use Algorithm \ref{Algo:noncooperate} to repetitively adjust their sharing decisions, $S_i$, at each round of the game in order to maximize its utility until reaching an equilibrium state (see Sect.~\ref{Sec:non_cooperative}); thus, the market-based model's computational time depends on the \emph{Tabu Search} distance and the number of SCs.
We consider scenarios with $2-8$ SCs in the federation, each with $100$ VMs. 
The number of iterations required decreases as more SCs participate (see Fig. \ref{Fig:time_complex_game}). This occurs because any decision change results in a bigger influence in a smaller federation. Similarly, a larger search distance's influence is bigger in a smaller federation. For example, our proposed market-based model needs $\approx 5$ iterations to reach equilibrium when only 2 SCs are in the federation.}

%Our extensive experimental evaluations indicate that if at least one SC chooses ${UF}_0$ as its utility function, the value of $\frac{C_{i}^{G}}{C_{i}^{P}}$ to maximize utilitarian SW is always greater than $0.7$; however, higher values might also increase the risk of breaking the cooperative, which depends on the utilization of SCs in the cooperative.
%
%However, when all SCs choose ${UF}_1$ as their utility function, the value of $\frac{C_{i}^{G}}{C_{i}^{P}}$ should be set to zero to maximize the utilitarian SW happens.
%
%If the goal of the cooperative is to achieve max-min fairness, the value of $\frac{C_{i}^{G}}{C_{i}^{P}}$ should be set between $0.3$ and $0.7$ in order to achieve maximum max-min SW (the real value depends the SCs' utilizations and utility functions).
%
%However, if proportional fairness is the goal, the value of $\frac{C_{i}^{G}}{C_{i}^{P}}$ should be set $0$ and $0.3$ when at least one SC chooses ${UF}_0$ as its utility function.
%
%Lastly, $\frac{C_{i}^{G}}{C_{i}^{P}} = 1$ can also incentive SCs to participate in the cooperative, particularly when all SCs have similar high utilizations.
% -*- ispell-local-dictionary: "american"; TeX-master: "cloudsharing.tex"; -*-
%
\section{Related Work}
We give an overview of efforts related to ours and highlight the relevant differences.
Works on hybrid clouds \cite{zhang2014proactive, shifrin2013optimal} are related
as they allow private (or smaller-scale) clouds
to outsource their requests to large-scale public providers. However, since that can potentially be costly
for a small-scale provider, our work differs in that it focuses on a sharing framework, while minimizing cost of using public clouds.

\sunghan{Earlier efforts also study the competition and cooperation within a federated cloud.
For instance, authors in \cite{goiri2012economic,hadji2015mathematical} characterize the cloud federation to help cloud providers maximize their profits via dynamic pricing models.
Earlier efforts \cite{truong2014novel,chen2017workload} also study the competition and cooperation among cloud providers, but assume that each cloud provider has sufficient resources to serve all users' requests, while \cite{chen2017workload} incorporates a penalty function to address the service delay penalty.
Authors in \cite{niyato2011resource} propose a hierarchical cooperative game theoretic model for better resources integration and achieving a higher profit in the federation.
Similar to our work, \cite{mashayekhy2015cloud} studies a federation formation game but assumes that cloud providers share everything with others, while \cite{hassan2014cooperative} adopts cooperative game theoretic approaches to model a cloud federation and study the motivation for cloud providers to participate in a federation.}

\sunghan{Another line of work focuses on designing sharing policies in the federation to obtain higher profit.
For instance, \cite{wangenabling} proposes a decentralized cloud platform \emph{SpotCloud} \cite{spotcloud} - a real-world system allowing customers or SCs to sell idle compute resources at specified prices -
and presents a resource pricing scheme (resulting from a repeated seller game) plus an optimal resource provisioning algorithm.
\cite{zhuang2014decentralizing} employs various cooperation strategies under varying workloads, to reduce the request rejection rate (i.e., the efficiency metric in \cite{zhuang2014decentralizing}).
Another effort \cite{toosi2011resource} trade-off the approaches of outsourcing resources and rejecting less profitable in order to increase resource utilization and profit.
\cite{wen2016cost} proposes to efficiently deploy distributed applications on federated clouds by considering security requirements, the cost of computing power, data storage and inter-cloud communication.
\cite{babaoglu2012design} groups resources of various SCs into computational units, in order to serve customers' requests.
\cite{samaan2014novel} proposes to incorporate both historical and expected future revenue into VM sharing decisions in order to maximize an SC's profit.}
%

%%%\emph{However, most of these efforts do not study the potential performance degradation received by each SC through participating in the cooperative, which is a significant factor for a SC to participate in the cooperative or not to; instead, they study the performance achievement of the whole cooperative.
%
%%%In contrast, our proposed effort is focused on achieving required
%%%performance characteristics of each SC while incentivizing them to participate in the cooperative, where
%%%some might be heavily loaded and others might have excess resources.

\noindent\textbf{Differences and Drawbacks.}
Our work differs from previous efforts in that we explicitly consider consequences of resource sharing on the resulting performance delivered to customers. 
In contrast, none of the above efforts explicitly model the system performance under the considered resource sharing environment. 
They either assume that resources can be reclaimed (when needed), thus resulting in lack of reliability of shared resources or they assume that an analytical performance characterization is possible (but do not propose a solution to estimate it).
Such an analytical characterization is an important contribution of our work.
To the best of our knowledge, this is the first work addressing the explicit interactions between performance model and economic model. 
Moreover, unlike previous efforts, that adopt the cooperative game theoretic approach, our work studying the non-cooperative game is more practical since likely no SC would be willing to share their utility specific information with others.

%assume the existence of the cloud federation and largely focus on designing sharing policies in order to maximize the profit of individual SCs, (2) most of them do not consider the risk of resource under-provision due to sharing too much resources - \emph{we focus on studying economic trade-off between the potential benefits (in terms of profit) and cost (in terms of performance degradation) for individual SCs, which are significant contributing factors in incentivizing SCs to participate in a cooperative}.
%

\section{Discussion and Future Work} \label{Sec:discussion}

%\sunghan{In this section, we discuss the rationale behind the assumptions made in the performance model and the market model.}

We made a number of assumptions in our models; here, we discuss the rationale behind the main assumptions.

\smallskip\noindent\textbf{Homogeneous VMs.} In practice, each cloud provider offers heterogeneous VM profiles (e.g., memory-optimized, CPU-optimized, or GPU-enabled), which reserve hardware resources on pre-specified machine pools shared by multiple VMs \cite{awstype}. However, many cloud providers, such as Amazon LightSail, DigitalOcean, and Linode, offer VM configurations with very similar specifications (e.g., \$10/month instances from Linode, DigitalOcean, and Amazon Lightsail currently provide 1 CPU core, 30 GB SSD, 2 TB data transfer/month, 1 or 2 GB of RAM). We believe that it is very likely that SCs would negotiate the sharing policies for each VM profile separately, given that these profiles correspond to different prices and capacities at each SC. In this case, our model of homogeneous resources can be applied repeatedly to each VM profile. Sharing policies for hardware resources (rather than VM profiles) would require the introduction of scheduling and packing algorithms within our performance model, which is beyond the scope of this work.

\smallskip\noindent\textbf{I.I.D. Exponential Service Times.} Depending on the target application, requests can require two or more VMs to complete a job, and service times of different requests likely have different distributions. In these cases, our Markov model can address non-exponential service times by introducing phase-type distributions that fit the moments of service time distributions from real-world traces \cite{osogami2006closed}. Similarly, batch arrivals can introduce with batch Markovian arrival processes (BMAPs). Unfortunately, both approaches result in larger state spaces, with the effect of increasing computation costs for the analysis of our performance model. In this paper, we motivated the formation of a federation using exponentially-distributed service times and single-VM requests to reduce the computational cost. To relax these assumptions, one of our future goals focuses on leveraging symbolic analysis methods for Markov chains, e.g., methods based on multi-terminal binary decision diagrams (MTBDDs), or lumping of Markov processes, to further cope with the state-space explosion.

\smallskip\noindent\textbf{Stable System Parameters.} This work focuses on establishing a long-term relationship in the federation: in reality, unlike spot clouds, where decisions must be made in a very short period of time, each SC would collect sufficient historical traces for a longer period of time before joining the federation, and update its sharing decisions after observing a long-term change in system parameters. Our approximate model is designed to deliver the results for this kinds of updates.

\noindent\textbf{Participating in single federation.} In real world, an SC can participate in multiple cloud federations simultaneously, and that sharing decisions for different federations might depend on many factors, such as the cost of using shared VMs in federations. However, this paper focuses on studying how the price of using shared resources affects the motivation of participating in the federation; profit maximization through the use of the resources from multiple federations is outside of the scope of this work, but it could represent future work.

\noindent\textbf{The feasibility of T\^{a}tonnement  process.} According to \cite{crawford2015essays}, when mixed strategies are considered, the results of the T\^{a}tonnement process might be unstable, which does
not happen with pure strategies. This entails (as one of the reasons) the use of pure strategies in practical settings. However, not all
games will have pure strategy equilibria, but they will definitely
have mixed strategy equilibria. In such situations, the T\^{a}tonnement
process to reach a pure strategy equilibria will not terminate,
indicating the possibility of the non-existence of a pure strategy
NE. Currently, we do not have a good solution to overcome this
problem. In addition, in all of our settings we reach a pure strategy
NE. Given the assumption that we only deal with pure strategies in our
game, the results from the T\^{a}tonnement process depend on the initial
point, particularly when there are more equilibria in the game. Thus,
in our game, we have tried different initial points, and picked the
equilibrium that produced a better fairness level among
SCs. Throughout our experiments, we can always find an equilibrium in
the game. However, as discussed in the previous comment, the existence
of an equilibrium significantly depends on the utilities of SCs.

\noindent\textbf{SCs follow the sequence of actions.} We stress the fact that it is in the rational
interest of users to follow the sequence/order as specified by the game. However, in the worst case, even if
users deviate from following the prescription, it is very unlikely that all users would do that at the same time.
In the event that even a few players follow the sequence/order as specified by the game, we would end up
with a better outcome than no-sharing. On an individual level, we agree with the reviewer that some players
might end up having a worse outcome than the scenario with no-sharing if they do not make new decisions
(e.g., leave the federation) when they have to sustain higher costs than in the case of not participating to
the federation. However, each SC can have a better utility than in the scenario with no sharing as long
as its decision can reduce the cost of serving customers, even when this SC does not constantly update its
decision.

\noindent\textbf{No collusion among SCs.} It is possible in
practice for certain SCs to collude among themselves to `game' the
cloud sharing system so that these SCs benefit more in terms of
resource availability at cheaper costs, than the others. It is here
that the benefits of a federation should come into play in two
possible ways: (a) enforce a strict set of laws prohibiting collusion,
in addition to strong punishments (e.g., being excluded from the
federation) if SCs are found to collude, and (b) designing economic
mechanisms (via the use of mechanism design models) to incentivize SCs
not to collude. However, the goal of this work focuses on studying how
the price of using shared resources will affect the decisions of SCs
that participate in the federation, and not on modeling collusions. We
leave the latter for future work.

\noindent\textbf{The same family of cost and utility function.} in practice different SCs might have cost and utility functions coming from different mathematical families. However, our design choice (to assume functions from the same family) is motivated by two practical insights stemming from our work:
\begin{itemize}
	\item One major element in our work focuses on discussing \emph{how the cost of resource usage affects the sharing strategies under different environments.} In this regard, we use the simple type of
	linear cost functions as a representative example of a cost
	function, which rationalizes realistic system designs (see Sect~\ref{Sec:cost}); in so doing, we reduce the complexity of our
	analysis. However, without loss of generality, other types of cost
	functions (even those coming from different mathematical families)
	using the same rationale as our work for the design of performance
	parameters, will show the same trend (albeit different values) when
	SCs change their sharing strategies: the reason is that different
	functions will exhibit similar mathematical properties of
	\emph{monotonicity}, \emph{continuity}, and
	\emph{differentiability}.
	
	\item A second important element in our work is the focus on \emph{studying fairness in sharing resources among SCs through the reduction of the cost (i.e,, $C_{i}^0 - C_{i}^{S_i}$ in Eq. (\ref{Eq:cost}))}. To this end, it is imperative that utility comparisons are done
		within a normalized interval range (e.g., [0,1]) irrespective of SCs
		having different families of utility functions in the worst
		case. This requires a formal normalization step which is outside the
		scope of our work. For simplicity, we assume that each SC utility is
		already normalized over a given fixed range: we implement this step
		in the experimental evaluation section by fixing the $\gamma$ value
		for each SC to be the same and varying between 0 and 1. For SC cost
		and utility functions from different mathematical families, after
		normalization, would produce similar trends and practical insights
		from fairness analysis, when compared to our experimental study.
\end{itemize}
	
	With respect to cost functions, we agree with the reviewer that the
	cost of using shared VMs from SCs might be different in the
	federation. However, our focus is on studying how the cost of using
	shared resources affects the motivation of participating in the cloud
	federation. If the cost of using shared resources is not homogeneous,
	the decision will also be affected by the resource allocation
	strategies (i.e., which SC to request the resources from). We do not
	try to introduce resource allocation strategies in this work, and thus
	assume that prices are homogeneous.
	In future work, we plan to study how the resource allocation
	strategies will affect the sharing decisions made by individual SCs.
	We also plan to incorporate different factors into our cost functions,
	such as trustfulness among SCs, and to propose a mechanism to evaluate
	multi-dimensional fairness for utility functions that belong to
	different mathematical families.

\smallskip\noindent\textbf{Future Work.} \emph{SC-Share} evaluates resource sharing benefits among SCs by accounting only for the cost of using VMs. However, there are other parameters that \emph{SC-Share} could account for in evaluating resource sharing benefits: (i) privacy concerns/risks of sharing/forwarding resources within cloud entities, (ii) data transmission costs for forwarding VM requests among cloud entities, and (iii) power consumption costs of running physical servers hosting VMs. We plan to incorporate these parameters into the \emph{SC-Share} framework as part of future work.

%These assumptions simplify our performance model, and where SCs allocate available VMs in the federation without preferences due to prices.

\section{Conclusions}

In this paper, we proposed \emph{SC-Share} for small-scale clouds (SCs) to enable them to share their resources in a profitable manner while maintaining customer SLAs.
Our framework is based on two (interacting) models: (i) an approximate performance model with an efficient solution that is able to produce sufficiently accurate estimates of performance characteristics of interest; and (ii) a market-based model that results in sharing policies which properly incentivize SCs to participate in the federation while achieving market success. 
%
%We illustrated the benefits of SC-Share numerically through an extensive performance and market evaluation study (SAY THE MAIN RESULT FROM THE SIMULATIONS)
%Through an extensive performance and market evaluation study numerically, 
\emph{SC-Share} can suggest different price settings in different federations in order to achieve sufficient market efficiency.
Moreover, \emph{SC-Share} shows that even when the price of shared VMs is equal to the price of using a public cloud, a federation can still be formed under certain criteria.

\bibliographystyle{IEEEtran}
\bibliography{IEEEabrv,cloudshare_arxiv}

\appendix

%\subsection{Conditional Probability Distribution} \label{Sec:conditional_probability}
%Given $X$, a discrete random variable over $\mathbb{N}=\{0,1, \dots\}$, and $\pi^X$, its probability mass function, the conditional probability distribution for subset $Y\subseteq X$ is:
%\begin{align}
%\pi^X(k) = \begin{cases}
% P_X(k | k\in Y) = \frac{P(X=k \wedge k\in Y)}{P(k\in Y)} & \text{if $ k\in Y$,}\\
% 0 & \text{otherwise.}
%\end{cases}
%\end{align}

\subsection{Mathematical Assumptions in Existing Theorems} \label{Sec:analyzing_market_detail}
Here, we describe in detail why existing seminal micro-economics theorems cannot be used to derive closed form results for market equilibria in our work. 

The seminal results by Nash provide a formal proof for both finite
(strategy spaces need not be continuous) and infinite (strategy space
is continuous) games that the existence of an equilibrium is possible
\cite{nash1950equilibrium} only when (i) the strategy set is compact,
i.e., closed and bounded, convex, and non-empty, and (ii) the utility
functions are necessarily quasi-concave (or stronger forms of
concavity) in a player's (mixed strategy) action, and continuous in
the vector of players actions. In addition, the theorem of Nash is
valid only for the guaranteed existence of a mixed strategy Nash
equilibrium. However, in our work we are only interested in the
guarantee of pure strategy Nash equilibria (see reason below), as it
is more practical to implement and realize. Thus, we design our SC's
utility function as a quasi-concave function. However, if SCs'
self-defined utility functions are non-quasi concave function, \emph{a
	mathematical proof for existing Nash equilibria in \emph{non-quasi
		concave} utility functions is still a difficult open problem based
	on Nash's theorem.}

In regard to guaranteeing a pure strategy Nash equilibria, consider
another seminal theorem by Debreu, Fan, and Glicksberg (derived
independently)
\cite{debreu1952social,glicksberg1952further,fan1952fixed} that
infinite games under the assumptions of (i) quasi-concavity of utility
functions, (ii) the utility functions being continuous in the vector
of players actions, and (iii) convex and compact strategy sets,
promise the existence of pure strategy Nash equilibrium. However, for
many practical settings including ours, the quasi-concavity assumption
might not always be satisfied (arbitrary SCs' utility function), which
in turn might not guarantee a pure strategy Nash equilibrium
(violating theorem assumptions). \emph{Thus, a mathematical proof for
	existing Nash equilibria in \emph{non-quasi concave} utility
	functions is still a difficult open problem based on the theorem by
	Glicksberg et.al.} In addition, strategy sets in many applications
(including specialized versions of our application setting, i.e., the
number of shared VMs is discrete in nature) might not be continuous,
in which case, we would have to go back to using Nash's theorem to
guarantee mixed strategy Nash equilibria.

An even stronger seminal theoretical result was proposed by Dasgupta
and Maskin \cite{dasgupta1986existence} that states: games under the
assumptions of (i) quasi-concavity of utility functions, (ii) the
utility functions being discontinuous (if we are dealing with
arbitrary utility functions as SCs might demand) in the vector of
players actions, and (iii) convex and compact strategy sets, promise
the existence of a mixed strategy Nash equilibrium. However, in our
work we are only interested in pure strategy Nash equilibria (see
reason below). \emph{Thus, a mathematical proof for existing Nash
	equilibria in \emph{non-quasi concave} utility functions is still a
	difficult open problem based on the theorem by Dasgupta and Maskin.}

Thus, we observe that practical modeling of a system might not always
fit the theoretical assumptions required to mathematically prove the
existence of a pure strategy Nash equilibria. \emph{Therefore, we
	resort to a simulation evaluation to search for the existence of
	Nash equilibria.} However, through simulation results, we do observe
the existence of pure strategy Nash equilibria for infinite strategy
spaces (simulated in a discrete manner, thereby becoming a finite game
in simulation), and for \emph{non quasi-concave} peer net utility
functions. Thus, at least from the experimental results, we observe
that for our work, (i) it is not necessary (via the theorem of Nash)
for quasi concavity to hold for a pure strategy (also discounting the
guarantee of only a mixed strategy via Nash's theorem) Nash
equilibrium to exist, and (ii) it is not necessary (via the theorem of
Debreu et.al.) for quasi concavity to hold for a pure strategy (also
discounting the infinite strategy space assumption via the theorem by
Debreu. et.al, as the simulation is discrete in nature) Nash
equilibrium to exist.  Taking all the above-mentioned issues in our
work related to fitting the assumptions required to prove the
existence of Nash equilibrium in theory, and the information
structure, we adopt a typical approach of \emph{fictitious play},
i.e., a time-averaged technique \cite{brown1951iterative} from the
theory of \emph{learning in games}, which allows us to converge upon a
Nash equilibria (provided its existence). However, we cannot guarantee
to reach the equilibrium if the SCs' self-defined utility functions
are non-quasi concave.

Finally, we report an historical perspective on the rationale for
studying pure strategy Nash equilibria:

\begin{displayquote}
	During the 1980s, the concept of mixed strategies came under heavy
	fire for being intuitively ``problematic.'' Randomization, central in
	mixed strategies, lacks behavioral support. Seldom do people make
	their choices following a lottery. This behavioral problem is
	compounded by the cognitive difficulty that people are unable to
	generate random outcomes without the aid of a random or pseudo-random
	generator. In 1991, game theorist Ariel Rubinstein described
	alternative ways of understanding the concept. The first, due to
	Harsanyi (1973), is called purification, and supposes that the mixed
	strategies interpretation merely reflects our lack of knowledge of the
	players' information and decision-making process. Apparently random
	choices are then seen as consequences of non-specified,
	payoff-irrelevant exogenous factors. However, it is unsatisfying to
	have results that hang on unspecified factors. Later, Aumann and
	Brandenburger (1995) \cite{brown1951iterative} re-interpreted Nash
	equilibrium as an equilibrium in beliefs, rather than actions. For
	instance, in the ``rock-paper-scissors'' game an equilibrium in
	beliefs would have each player believing the other was equally likely
	to play each strategy. This interpretation weakens the predictive
	power of Nash equilibrium, however, since it is possible in such an
	equilibrium for each player to actually play a pure strategy of
	Rock. Ever since, game theorists' attitude towards mixed
	strategies-based results have been ambivalent. Mixed strategies are
	still widely used for their capacity to provide Nash equilibria in
	games where no equilibrium in pure strategies exist, but the model
	does not specify why and how players randomize their decisions.
	\flushright From: Strategy (game theory), \emph{Wikipedia}
\end{displayquote}

\end{document}